\documentclass{sig-alternate}
\usepackage{multirow}
\usepackage{subfig}
\usepackage{graphicx}

\newcounter{subcopyrightbox@save}
\usepackage{caption}
\usepackage{color, url}
\usepackage{xspace}

\newcommand{\neil}[1]{{\small\color{blue}[Neil: #1]}}

\newcommand{\eat}[1]{}

\newcommand{\gplus}{Google{\small $+$}\xspace}
\newtheorem{theorem}{Theorem}
\newtheorem{definition}[theorem]{Definition}

\begin{document}

\conferenceinfo{The 6th SNA-KDD Workshop(SNA-KDD'12),}{Aug.12, 2012, Beijing, China}
\CopyrightYear{2012} % Allows default copyright year (20XX) to be over-ridden - IF NEED BE.
\crdata{978-1-4503-1544-9}  % Allows default copyright data (0-89791-88-6/97/05) to be over-ridden - IF NEED BE.
\title{Jointly Predicting Links and Inferring Attributes using a Social-Attribute Network (SAN)}

%\numberofauthors{8} 

\author{
\hspace{1mm} Neil Zhenqiang Gong \\
\hspace{1mm} \url{neilz.gong@berkeley.edu}
\and
Ameet Talwalkar\\
\url{ameet@cs.berkeley.edu}
\and
\hspace{-4mm} Lester Mackey \\
\hspace{-4mm} \url{lmackey@cs.berkeley.edu }
\and
\hspace{-4mm} Ling Huang \\
\hspace{-4mm} \url{ling.huang@intel.com }
\and  
\hspace{-2mm} Eui Chul Richard Shin \\
\hspace{-2mm} \url{ricshin@berkeley.edu}
\and
\hspace{-1mm} Emil Stefanov             \\
\hspace{-1mm} \url{emil@cs.berkeley.edu}
\and
\hspace{-1mm} Elaine (Runting) Shi             \\
\hspace{-1mm} \url{elaines@cs.berkeley.edu}
\and
\hspace{-7mm} Dawn Song \\
\hspace{-7mm} \url{dawnsong@cs.berkeley.edu}}

\maketitle

\begin{abstract}
The effects of social influence and homophily suggest that both network
structure and node attribute information should inform the tasks of link
prediction and node attribute inference. Recently, Yin et al.~\cite{Yin10-1,
Yin10} proposed \emph{Social-Attribute Network} (SAN), an attribute-augmented
social network, to integrate
network structure and node attributes to perform both link prediction and
attribute inference. They focused on generalizing the random walk with restart
algorithm to the SAN framework and showed improved performance.  
In this paper, we extend the SAN framework with several leading supervised and
unsupervised link prediction algorithms and demonstrate performance improvement
for each algorithm on both link prediction and attribute inference.
Moreover, we make the novel observation that attribute
inference can help inform link prediction, i.e., link prediction accuracy is
further improved by first inferring missing attributes. 
We comprehensively evaluate these algorithms and compare them with 
other existing algorithms using a novel, large-scale \gplus dataset, 
which we make publicly available\footnote{\scriptsize\url{http://www.cs.berkeley.edu/~stevgong/dataset/snakdd12.zip}}.

\end{abstract}
 
\category{H.4}{Information Systems Applications}{Miscellaneous}

\terms{Social Network}

\keywords{Link prediction, Attribute inference, Social-Attribute Network (SAN)}

\section{Introduction}

Online social networks (e.g., Facebook, \gplus) have become
increasingly important resources for interacting with people, processing
information and diffusing social influence.  
%These networks are also highly dynamic.  
Understanding and modeling the mechanisms by which these networks evolve
are therefore fundamental issues and active areas of research.

The classical \emph{link prediction problem} \cite{link-pre-survey03} has
attracted particular interest. In this setting, we are given a snapshot of a
social network at time $t$ and aim to predict links (e.g., friendships) that
will emerge in the network between $t$ and a later time $t'$.  Alternatively, we can
imagine the setting in which some links existed at time $t$ but are missing at
$t'$. In online social networks, a change in privacy settings often leads to
missing links, e.g., a user on \gplus might decide to hide her family circle
between time $t$ and $t'$. The missing link problem has important ramifications
as missing links can alter estimates of network-level
statistics~\cite{infer-missing-link06}, and  the ability to infer these missing
links raises serious privacy concerns for social networks.  Since the same
algorithms can be used to predict new links and missing links, we refer to
these problems jointly as link prediction. 

Another problem of increasing interest revolves around node attributes~\cite{Zheleva09}. Many
real-world networks contain rich categorical node attributes, e.g., users in
\gplus have profiles with attributes including employer, school, occupation and
places lived.  In the \emph{attribute inference problem}, we aim to
populate attribute information for network nodes with missing or incomplete
attribute data. This scenario often arises in practice when users in online
social networks set their profiles to be publicly invisible or create an
account without providing any attribute information.  The growing interest in
this problem is highlighted by the privacy implications associated with
attribute inference as well as the importance of attribute information for
applications including people search and collaborative filtering.
%applications including people search \cite{people-search-02} and collaborative
%filtering \cite{collaborativefiltering10}.

%============== old version ===================
\eat{
\neil{we say network structure and node attributes are entangled. }
In this work, we simultaneously use network structure and node attribute
information to improve performance on the link prediction and the attribute
inference problems. The principle of homophily
\cite{homopily04,homopily01,homopily11}, which states that users with similar
attributes are likely to link to one another, motivates the use of attributes
for link prediction.  Similarly, the principle of social influence
\cite{homopily11}, which states that users who are linked are likely to adopt
similar attributes, suggests that network structure should inform attribute
inference.  Additionally, previous studies \cite{Watts06,homopily11} have
empirically demonstrated the effects of homophily and social influence on
real-world social networks, providing further support for considering both
network structure and node attribute information when predicting links or
inferring attributes. 
}
%================================================

%\ling{changed homophily in this paragraph.}
In this work, we simultaneously use network structure and node 
attribute information to improve performance of both the link prediction 
and the attribute inference problems, motivated by the observed
interaction and homophily between network structure and node 
attributes. % \cite{homopily04, Watts06, Zheleva09, homopily11, MAG11}. 
%\lester{Removed these citations to save space.  We cite them again below.}
The principle of social influence \cite{homopily11}, which states 
that users who are linked are likely to adopt similar attributes, 
suggests that network structure should inform attribute inference.  Other
evidence of interaction \cite{homopily04, MAG11} shows that users with
similar attributes, or in some cases antithetical attributes, are likely to
link to one another, motivating the use of attribute information for link
prediction.  Additionally, previous studies \cite{Watts06,homopily11} have
empirically demonstrated those effects on real-world social networks, providing
further support for considering both network structure and node attribute
information when predicting links or inferring attributes.

However, the algorithmic question of how to simultaneously incorporate these
two sources of information remains largely unanswered. 
%
%Link prediction methods
%that aim to leverage attribute information have appeared in the relational
%learning community \cite{relational-link-pre03, latent-link-pre09, Yu06} and matrix factorization and alignment community~\cite{matrix_factorization, matrix_alignment}, 
%
The relational learning~\cite{relational-link-pre03, latent-link-pre09, Yu06}, matrix factorization and alignment~~\cite{matrix_factorization, matrix_alignment} based approaches
have been proposed to leverage attribute information for link prediction,
%that aim to leverage attribute information have appeared in the relational
%learning community \cite{relational-link-pre03, latent-link-pre09, Yu06} and matrix factorization and alignment community~\cite{matrix_factorization, matrix_alignment}, 
but they suffer from scalability issues. 
%
%Recent approaches based on matrix factorization~\cite{matrix_factorization} 
%and alignment~\cite{matrix_alignment}
%show superio performance over many algorithms, but they suffer from high
%computation complexity.
%
More recently, Backstrom and Leskovec~\cite{srw11} presented a
Supervised Random Walk (SRW) algorithm for link prediction that combines
network structure and edge attribute information, but this approach does not
fully leverage node attribute information as it only incorporates node
information for neighboring nodes. For instance, SRW cannot take advantage of
the common node attribute San Francisco of $u_2$ and $u_5$ in
Fig.~\ref{figure:san} since there is no edge between them. 

Yin et al.~\cite{Yin10, Yin10-1} proposed the use of \emph{Social-Attribute
Network} (SAN) to gracefully integrate network structure and node 
attributes in a
scalable way. They focused on generalizing Random Walk with Restart (RWwR)
algorithm to the SAN model to predict links as well as infer node attributes.
In this paper,  we generalize several leading supervised
and unsupervised link prediction algorithms \cite{link-pre-survey03,Hasan06} to
the SAN model to both predict links and infer missing attributes. We
evaluate these algorithms on a novel, large-scale \gplus dataset,  and 
demonstrate performance improvement for each of them.  
Moreover, we make the novel observation that
inferring attributes could help predict links, i.e., link prediction accuracy
is further improved by first inferring missing node attributes. 

%
%The rest of the paper is organized as follows: we formally define our problem
%in Section 2, introduce the SAN model and generalized algorithms for link
%prediction and attribute inference in Section 3, describe data collection and
%preprocessing in Section 4, discuss the experimental results in Section 5,
%review related work in Section 6 and conclude the paper with future work in
%Section 7.

\section{Problem Definition}

In our problem setting, we use an undirected\footnote{\scriptsize Our model and algorithms
can also be generalized to directed graphs.} graph $G=(V, E)$ to represent a
social network,  where edges in $E$ represent interactions between the $N =
|V|$ nodes in $V$.  In addition to network structure, we have categorical
attributes for nodes. For instance, in the \gplus social network, nodes are
users, edges represent friendship (or some other relationship) between users,
and node attributes are derived from user profile information and include
fields such as employer, school, and hometown. In this work we restrict our
focus to categorical variables, though in principle other types of variables,
e.g., live chats, email messages, real-valued variables, etc., could be
clustered into categorical variables via vector quantization, or directly
discretized to categorical variables.

We use a binary representation for each categorical attribute. For example,
various employers (e.g., Google, Intel and Yahoo) and various schools (e.g.,
Berkeley, Stanford and Yale) are each treated as separate binary attributes.
Hence, for a specific social network, the number of distinct attributes $M$ is
finite (though $M$ could be large). Attributes of a
node $u$ are then represented as a $M$-dimensional trinary column vector
$\vec{a}_u$ with the $i^{th}$ entry equal to $1$ when $u$ has the $i^{th}$
attribute (\emph{positive attribute}), $-1$ when $u$ does not have it
(\emph{negative attribute}) and $0$ when it is unknown whether or not $u$ has
it (\emph{missing attribute}). We denote by $A=[\vec{a}_1\ \vec{a}_2 \cdot
\cdot \cdot \vec{a}_N]$ the attribute matrix for all nodes. Note that
certain attributes (e.g. Female and Male, age of 20 and 30) are mutually exclusive. Let $L$ be the
set of all pairs of mutually exclusive attributes.  This set constrains the 
attribute matrix $A$ so that no column contains a $1$ for two mutually exclusive attributes.  

We define the link prediction problem as follows:

\begin{definition}[Link Prediction Problem]
Let $T_i=(G_i,A_i,L_i)$ and $T_j=(G_j,A_j,L_j)$ be snapshots of a social network at
times $i$ and $j$.  Then the link prediction problem involves using $T_i$ to
predict the social network structure $G_j$. When $i<j$, new links are predicted.
When  $i>j$, missing links are predicted.
\end{definition}

In this paper, we work with three snapshots of the \gplus network crawled at
three successive times, denoted $T_1=(G_1,A_1,L_1)$, $T_2=(G_2,A_2,L_2)$ and
$T_3=(G_3,A_3,L_3)$. To predict new links, we use various algorithms to solve the
link prediction problem with $i=2$ and $j=3$ and first learn any required
hyperparameters by performing grid search on the link prediction
problem with $i=1$ and $j=2$.  Similarly, to predict missing links, we solve the
link prediction problem with $i=2$ and $j=1$ and learn hyperparameters via
grid search with $i=3$ and $j=2$.  

For any given snapshot, several entries of $A$ will be zero, corresponding to missing attributes. The attribute inference problem, which involves only a single snapshot of the network, is
defined as follows:

\begin{definition}[Attribute Inference Problem]
Let $T=(G,A,L)$ be a snapshot of a social network. Then the attribute inference
problem is to infer whether each zero entry of $A$
corresponds to a positive or negative attribute, subject to the constraints listed in $L$.
\end{definition}

Our goal is to design scalable algorithms leveraging both network structure and
rich node attributes to address these problems for real-world large-scale
networks.

\section{Model and Algorithms}
\subsection{Social-Attribute Network Model}
\emph{Social-Attribute Network} was first proposed by Yin et al.~\cite{Yin10-1, Yin10}\footnote{\scriptsize Note that they name this model as \emph{Augmented Graph}. We call it as \emph{Social-Attribute Network} because it's more meaningful.} to predict links and infer attributes. However, their original model didn't consider negative and mutually exclusive attributes. In this section, we review this model and extend it to incorporate negative and mutex attributes.
 
Given a social network $G$ with $M$ distinct categorical attributes, an
attribute matrix $A$ and mutex attributes set $L$, we create an augmented
network by adding $M$ additional nodes to $G$, with each additional node
corresponding to an attribute.  For each node $u$ in $G$ with positive or
negative attribute $a$, we create an undirected link between $u$ and $a$ in the
augmented network. For each mutually exclusive attribute pair $(a,b)$, we
create an undirected link between $a$ and $b$. This augmented network
is called the \emph{Social-Attribute Network} (SAN) since it includes the original social
network interactions, relations between nodes and their attributes and mutex
links between attributes.

\begin{figure}
\centering
\includegraphics[width=0.4 \textwidth, height=1.6 in]{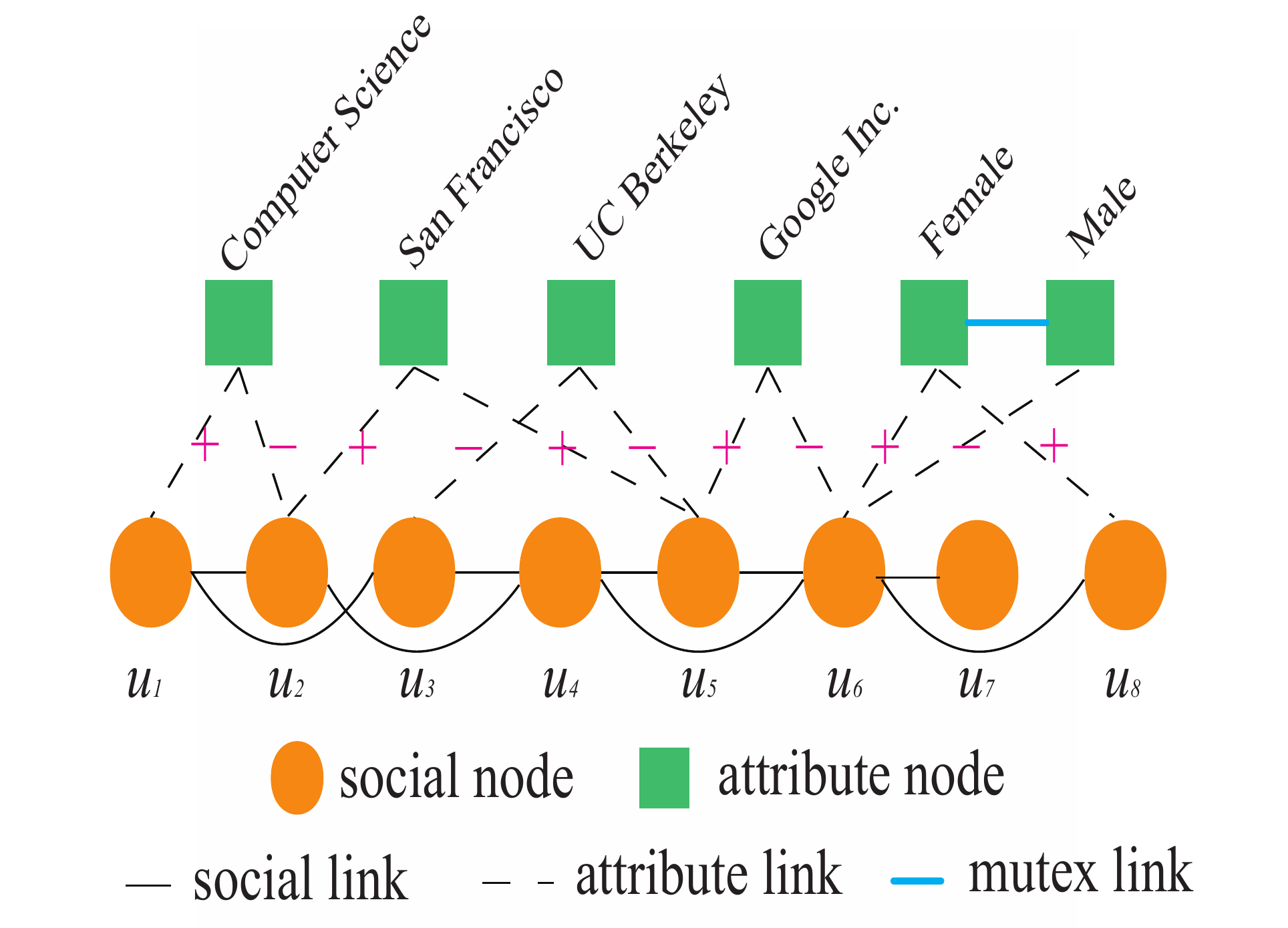}
\captionsetup{font=small,labelfont=bf}
\caption{\bf  Illustration of a Social-Attribute Network (SAN). The link prediction problem reduces to predicting
social links while the attribute inference problem involves predicting
attribute links.}
\label{figure:san}
\vspace{-4mm}
\end{figure}

Nodes in the SAN model corresponding to nodes in $G$ are called \emph{social
nodes}, and nodes representing attributes are called \emph{attribute nodes}.
Links between social nodes are called \emph{social links}, and links between
social nodes and attribute nodes are called \emph{attribute links}. Attribute
link $(u,a)$ is a \emph{positive attribute link} if $a$ is a positive attribute
of node $u$, and it is a \emph{negative attribute link} otherwise. Links between 
mutually exclusive attribute nodes are called \emph{mutex links}. Intuitively,
the SAN model explicitly describes the sharing of attributes across social
nodes as well as the mutual exclusion between attributes, as illustrated in the
sample SAN model of Fig.~\ref{figure:san}.  Moreover, with the SAN model, the
link prediction problem reduces to predicting social links and the attribute
inference problem involves predicting attribute links.

We also place weights on the various nodes and edges in the SAN model.  These
node and edge weights describe the relative importance of individual nodes or
relationships across nodes and can also be used in a global fashion to balance
the influence of social nodes versus attribute nodes and social links versus
attribute links. We use $w(u)$ and $w(u,v)$ to denote the weight of node $u$
and the weight of link $(u,v)$, respectively. 
Additionally, for a given social or attribute node $u$ in the SAN model, we denote by 
$\Gamma_+(u)$ and $\Gamma_{s+}(u)$ respectively the set of \emph{all neighbors} and 
the set of \emph{social neighbors} connected to $u$ via social links or positive attribute links.
We define $\Gamma_-(u)$ and $\Gamma_{s-}(u)$ in a
similar fashion. This terminology will prove useful when we describe our
generalization of leading link prediction algorithms to the SAN model in the
next section.

The fact that no social node can be linked to multiple mutex
attributes is encoded in the \emph{mutex property}, i.e., there is no triangle consisting of a mutex
link and two positive attribute links in any social-attribute network, which enforces a set of constraints for all attribute inference algorithms.

In this work, we focus primarily on node attributes.  However, we note that the
SAN model can be naturally extended to incorporate \emph{edge attributes}.
Indeed, we can use a function (e.g., the logistic function) to map a given set of 
attributes for each edge (e.g., edge age) into the real-valued edge weights of the SAN model.
The attributes-to-weight mapping function can be learned using an
approach similar to the one proposed by Backstrom and Leskovec~\cite{srw11}.  

\subsection{Algorithms}
\label{sec:algorithms}

Link prediction algorithms typically compute a probabilistic score 
for each candidate link and subsequently rank these scores and choose the largest ones (up to some
threshold) as putative new or missing links.  In the following, we extend both
unsupervised and supervised algorithms to the SAN model. % to compute the scores.
Furthermore, we note that when predicting attribute links, the SAN model 
features a post-processing step whereby we change the lowest ranked 
putative positive links violating the mutex property to negative links. 

\subsubsection{Unsupervised Link and Attribute Inference}
\label{unsupervise}
Liben-Nowell and Kleinberg \cite{link-pre-survey03} provide a comprehensive
survey of unsupervised link prediction algorithms for social networks. These algorithms can be roughly divided into two categories:
local-neighborhood-based algorithms and global-structure-based algorithms. In
principle, all of the algorithms discussed in \cite{link-pre-survey03} can be
generalized for the SAN model.  In this work we focus on representative
algorithms from both categories and we describe below how to generalize them to
the SAN model to predict both social links and attribute links. We add the
suffix `-SAN' to each algorithm name to indicate its generalization to the SAN
model. In our presentation of the unsupervised algorithms, we only consider positive attribute links, though
many of these algorithms can be extended to signed networks \cite{Symeonidis10}.  \\

\noindent
{\bf Common Neighbor (CN-SAN)} is a local algorithm that computes a score for a
candidate social or attribute link $(u,v)$ as the sum of weights of $u$ and
$v$'s common neighbors, i.e.  $score(u,v)=\sum_{t\in \Gamma_+(u) \cap \Gamma_+(v)}
w(t)$. Conventional CN only considers common social neighbors. \\

\noindent
{\bf Adamic-Adar (AA-SAN)} is also a local algorithm.  For a candidate
social link $(u,v)$ the AA-SAN score is $$score(u,v)=\sum_{t\in \Gamma_+(u) \cap
\Gamma_+(v)} \frac{w(t)}{\log |\Gamma_{s+}(t)|}\,.$$  Conventional AA, initially
proposed in~\cite{aa03} to predict friendships on the web and subsequently
adapted by \cite{link-pre-survey03} to predict links in social networks, only
considers common social neighbors.  AA-SAN weights the importance of a common
neighbor proportional to the inverse of the log of social degree.  Intuitively,
we want to downweight the importance of neighbors that are either i) social
nodes that are social hubs or ii) attribute nodes corresponding to attributes
that are widespread across social nodes.  Since in both cases this weight
depends on the social degree of a neighbor, the AA-SAN weight is derived based
on social degree, rather than total degree.  

In contrast, for a candidate attribute link $(u,a)$, the attribute degree of a
common neighbor does influence the importance of the neighbor. 
%(since attribute nodes have no attribute links, this argument pertains only to social nodes).  
%\lester{Attributes have mutex links. I don't think it's worth the space to make this statement precise.}
For instance, consider two social nodes with the same social degree that are
both common neighbors of nodes $u$ and $a$.  If the first of these social nodes
has only two attribute neighbors while the second has $1000$ attribute
neighbors, the importance of the former social node should be greater with
respect to the candidate attribute link.  Thus, AA-SAN computes the score for
candidate attribute link $(u,a)$ as
$$score(u,a)=\sum_{t\in \Gamma_{s+}(u) \cap \Gamma_{s+}(a)} \frac{w(t)}{\log |\Gamma_+(t)|}
\,.$$ 

\noindent
{\bf Low-rank Approximation (LRA-SAN)} takes advantage of global structure,
in contrast to CN-SAN and AA-SAN. Denote $X_S$ as the $N\times N$
weighted social adjacency matrix where the $(u,v)th$ entry of $X_S$ is $w(u,v)$
if $(u,v)$ is a social link and zero otherwise.  Similarly, let $X_A$ be the $N
\times M$ weighted attribute adjacency matrix where the $(u,a)th$ entry of
$X_A$ is $w(u,a)$ if $(u,a)$ is a positive attribute link and zero otherwise.  We then
obtain the weighted adjacency matrix $X$ for the SAN model by concatenating
$X_S$ and $X_A$, i.e., $X=[X_S \ X_A]$. The LRA-SAN method assumes that a small
number of latent factors (approximately) describe the social and attribute link
strengths within $X$ and attempts to extract these factors via low-rank
approximation of $X$, denoted by $\hat X$.  The LRA-SAN score for a candidate
social or attribute link $(u,t)$ is then simply  $\hat X_{ut}$, or the
$(u,t)th$ entry of $\hat X$.  LRA-SAN can be computed efficiently via truncated
Singular Value Decomposition (SVD).\\

\noindent
{\bf CN + Low-rank Approximation (CN+LRA-SAN)} is a mixture of
local and global methods, as it first performs CN-SAN using a SAN model and
then performs low-rank approximation on the resulting score matrix. After
performing CN-SAN, let $S_S$ be the resulting $N\times N$ score matrix for all
social node pairs and $S_A$ be the resulting $N \times M$ score matrix for all
social-attribute node pairs. By virtue of the CN-SAN algorithm, note that $S_S$
includes attribute information and $S_A$ includes social interactions.
CN+LRA-SAN then predicts social links by computing a low-rank approximation of
$S_S$ denoted $\hat{S}_S$, and each entry of $\hat{S}_S$ is the predicted
social link score.  Similarly, $\hat{S}_A$ is a low-rank approximation of
$S_A$, and each entry of $\hat{S}_A$ is the predicted score for the
corresponding attribute link.\footnote{\scriptsize An alternative method for combining
CN-SAN and LRA-SAN under the SAN model that was not explored in this work
involves defining $S=[S_S \ S_A]$, approximating $S$ with $\hat{S}$ and using
the $(u,t)th$ entry of $\hat{S}$ as a score for link $(u,t)$.} \\

\noindent
{\bf AA + low-rank Approximation(AA+LRA-SAN)} is identical to
CN+LRA-SAN but with the score matrices $S_S$ and $S_A$ generated via the AA-SAN
algorithm. \\

\noindent
{\bf Random Walk with Restart (RWwR-SAN)}~\cite{Yin10} is a global algorithm.
In the SAN model, a Random Walk with Restart ~\cite{pagerank98, RWwR03}
starting from $u$ recursively walks to one of its neighbors $t$ with
probability proportional to the link weight $w(u,t)$ and returns to $u$  with a
fixed restart probability $\alpha$.  The probability $P_{u,v}$ is the
stationary probability of node $v$ in a random walk with restart initiated at
$u$. In general, $P_{u,v}\neq P_{v,u}$. For a candidate social link $(u,v)$, we
compute $P_{u,v}$ and $P_{v,u}$ and let $score(u,v) = (P_{u,v} +P_{v,u})/2$.
Note that RWwR for link prediction in previous work~\cite{link-pre-survey03}
computes these stationary probabilities based only on the social network. For a
candidate attribute link $(u,a)$, RWwR-SAN only computes $P_{u,a}$, and
$P_{u,a}$ is taken as the score of $(u,a)$. \\

We finally note that for predicting social links, if we set the weights of all
attribute nodes and all attribute links to zero and we set the weights of all
social nodes and social links to one, then all the algorithms described above
reduce to their standard forms described
in~\cite{link-pre-survey03}.\footnote{\scriptsize For LRA-SAN this implies that $X_A$ is an
$N \times M$ matrix of zeros, so the truncated SVD of $X$ is
equivalent to that of $X_S$ except for $M$ zeros appended to the right singular
vectors of $X_S$.}  In other words, we recover the link prediction algorithms
on pure social networks.

\subsubsection{Supervised Link and Attribute Inference}
\label{supervise}
Link prediction can be cast as a binary classification problem, in which we first
construct features for links, and then use a classifier such as SVMs or
Logistic Regression. In contrast to unsupervised attribute inference, negative attribute links are needed in supervised attribute inference.\\

\noindent
{\bf Supervised Link Prediction (SLP-SAN)} 
%New or missing links are viewed as positive examples while non-existing links are viewed as negative examples.
%Each link is represented by a feature vector with features extracted from pure
%social networks or social-attribute networks or both. 
For each link in our training set, we can extract a set of topological features
$F$ (e.g. CN, AA, etc.) computed from pure social networks and the similar
features $F\_SAN$ computed from the corresponding social-attribute networks. We
explored 4 feature combinations: i) SLP-I uses only topological features $F$
computed from social networks; ii) SLP-II uses topological features $F$ as well
as an aggregate feature, i.e., the number of common attributes of the two
endpoints of a link; iii) SLP-SAN-III uses topological features $F\_SAN$; and
iv) SLP-SAN-VI uses topological features $F$ and $F\_SAN$.  SLP-SAN-III and
SLP-SAN-VI contain the substring `SAN'  because they use features extracted
from the SAN model. SLP-I and SLP-II are widely used in previous
work~\cite{Hasan06, Lichtenwalter10, srw11}.\\

\noindent
{\bf Supervised Attribute Inference (SAI-SAN)}  Recall that attribute inference
is transformed to attribute link prediction with the SAN model. We can extract
a set of topological features for each positive and negative attribute link.
Moreover, the positive attribute links are taken as positive examples while the
negative attribute links are taken as negative examples. Hence, we can train a
binary classifier for attribute links and then apply it to infer the missing
attribute links.

\subsubsection{Iterative Link and Attribute Inference}
\begin{figure}
\centering
\includegraphics[width=0.25 \textwidth, height=1.5in]{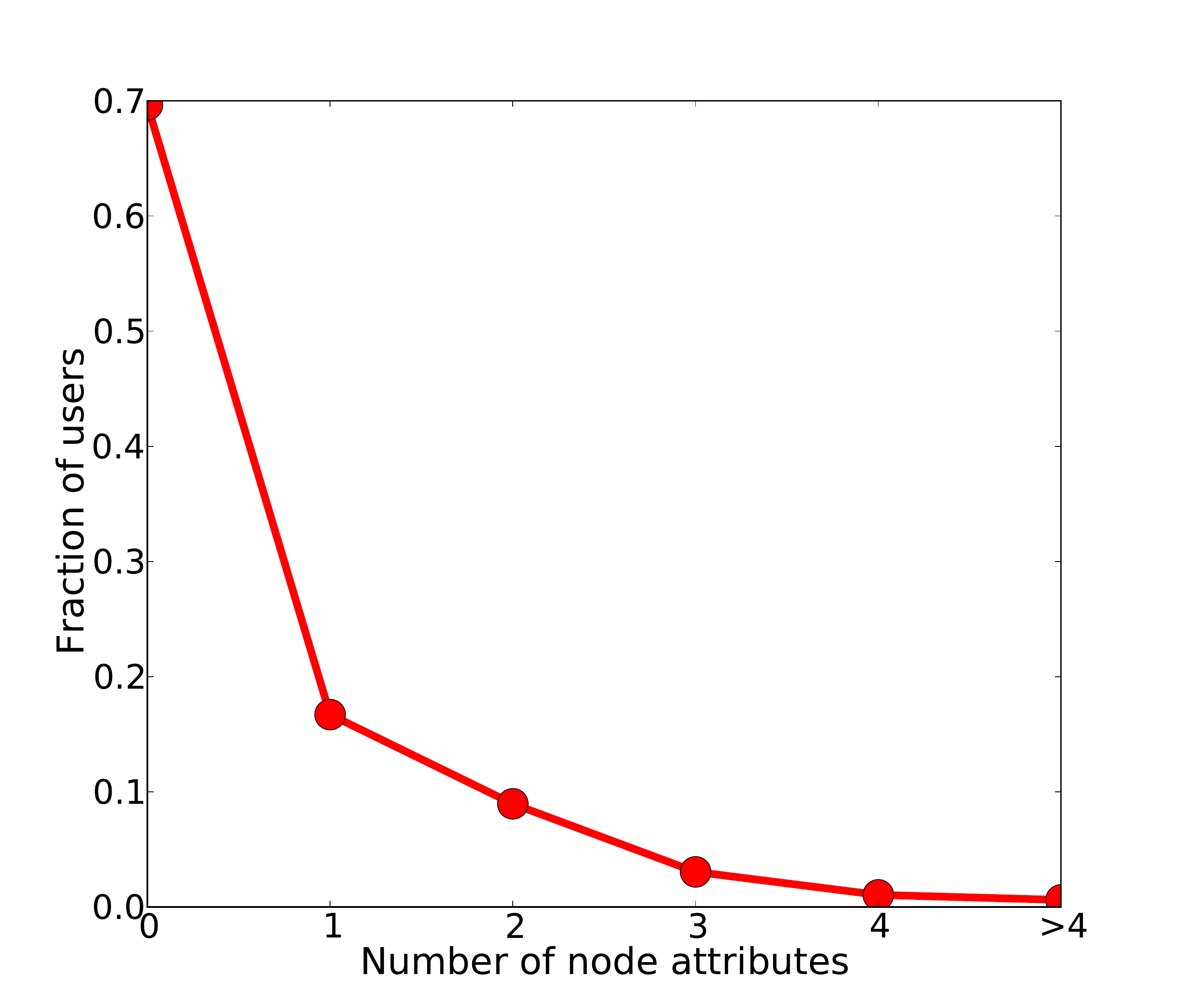}
\captionsetup{font=small,labelfont=bf}
\caption{\bf  The fraction of users as a function of the number of node
attributes in the \gplus social network. }
\label{figure:node-attri}
\vspace{-4mm}
\end{figure}

In many real-world networks, most node attributes are missing. Fig.
\ref{figure:node-attri} shows the fraction of users as a function of the number
of node attributes in \gplus social network. From this figure, we see that
roughly 70\% of users have no observed node attributes.  Hence, we will also
investigate an iterative variant of the SAN model.  We first infer the top
attributes for users without any observed attributes.  We then update the SAN model to
include these predicted attributes and perform link prediction on the
updated SAN model.  This process can be performed for several iterations.

\section{Google+ Data}

Google launched its new social network service named \gplus\ in early July
2011. We crawled three snapshots of the \gplus\ social network and their users'
profiles on July 19, August 6 and September 19 in 2011.  They are
denoted as JUL, AUG and SEP, respectively.  We then pre-processed the data
before conducting link prediction and attribute inference experiments. \\

\noindent
{\bf Preprocessing Social Networks} In \gplus, users divide their social
connections into circles, such as a family circle and a friends circle. If user
$u$ is in $v$'s circle, then there is a directed edge $(v,u)$ in the graph, and
thus the \gplus dataset is a directed social graph. We converted this dataset
into an undirected graph by only retaining edges $(u,v)$ if both directed edges
$(u,v)$ and $(v,u)$ exist in the original graph.  We chose to adopt this
filtering step for two reasons: (1) Bidirectional edges represent mutual
friendships and hence represent a stronger type of relationship that is more
likely to be useful when inferring users' attributes from their friends'
attributes (2) We reduce the influence of spammers who add people into their
circles without those people adding them back. Spammers introduce fictitious
directional edges into the social graph that adversely influence the
performance of link prediction algorithms.\\

\noindent
{\bf Collecting Attribute Vocabulary} \gplus profiles include short entries
about users such as Occupation, Employment, Education, Places
Lived, and Gender, etc. We use Employment and
Education to construct a vocabulary of attributes in this paper.  We treat
each distinct employer or school entity as a distinct attribute. \gplus\ has
predefined employer and school entities, although users can still fill in their
own defined entities. Due to users' changing privacy settings, some profiles in
JUL are not found in AUG and SEP, so we use JUL to construct our attribute
vocabulary.  Specifically, from the profiles in JUL, we list all
attributes and compute frequency of appearance for each attribute.  Our
attribute vocabulary is constructed by keeping attributes with frequency of at
least 3.\\

\noindent
{\bf Constructing Social-Attribute Networks} In order to demonstrate that the
SAN model leverages node attributes well, we derived social-attribute
networks in which each node has some positive attributes from the above \gplus
social networks and attribute vocabulary. Specifically, for an
attribute-frequency threshold $k$, we chose the largest connected social
network from JUL such that each node has at least $k$ distinct positive attributes. We
also found the corresponding social networks consisting of these nodes in
snapshots AUG and SEP. Social-attribute networks were then constructed with the
chosen social networks and the attributes of the nodes.  Specifically, we chose
$k= \{2, 4 \}$  to construct 6 social-attribute networks whose statistics are
shown in Table~\ref{table:net_stat}. Each social-attribute network is named by
concatenating the snapshot name and the attribute-frequency threshold. For
example, `JUL4' is the social-attribute network constructed using JUL and
$k=4$.  These names are indicated in the first column of the table.  

%Table \ref{table:net_stat} shows the statistics of the social-attribute networks.
In the crawled raw networks, some social links in JUL$i$ are missing in AUG$i$ and
SEP$i$, where $i=2,4$.  These links are missing due to one of two events
occurring between the JUL and AUG or SEP snapshots: 1) users block other users,
or 2) users set (part of) their circles to be publicly invisible after
which point they cannot be publicly crawled.  These missed links provide ground
truth labels for our experiments of predicting missing links. However, these
missing links can alter estimates of network-level statistics, and can have
unexpected influences on link prediction
algorithms~\cite{infer-missing-link06}.  Moreover, it is likely in practice
that companies like Facebook and Google keep records of these missing links,
and so it is reasonable to add these links back to AUG$i$ and SEP$i$ for our
link prediction experiments. The third column in Table \ref{table:net_stat} is the number of all social links after filling the missing links into AUG$i$ and SEP$i$.  The second column \emph{\#soci links} is used 
for experiments of predicting missing links, and column \emph{\#all soci links} is used for the experiments of predicting new links.

From these two columns, the number of new links or missing links can be easily
computed. For example, if we use AUG2 as training data and SEP2 as testing data
for link prediction, the number of new links is $354572-339059=15513$, which is
computed with entries in column \emph{\#all soci links}. If we use AUG2 as
training data and JUL2 as testing data in predicting missing links, the number
of missing links is $339059-328761=10298$, which is computed with corresponding
entries in column \emph{\#soci links} and \emph{\#all soci links}.

\begin{table}[t]\renewcommand{\arraystretch}{0.6}
\centering
\captionsetup{font=small,labelfont=bf}
\caption{\bf Statistics of social-attribute networks.}
%\caption{\bf Statistics of social-attribute networks. Column \emph {\#soci
%links} is the number of social links in the crawled raw datasets. Column
%\emph{\#all soci links} is the number of social links after filling in missing
%social links.}
%\subfloat[]{
%\hspace{-1.cm}
\centering
\addtolength{\tabcolsep}{-5pt}
\begin{tabular}{|c|c|c|c|c|c|} \hline
& {\tiny \#soci links} & {\tiny \#all soci links} & {\tiny \#soci nodes}& {\tiny\#pos attri links} &{\tiny\#attri nodes}\\ \hline

{\tiny JUL4} & {\tiny 7062}& {\tiny 7062} &\multirow{3}{*}{{\tiny 5200}} & \multirow{3}{*}{{\tiny 24690}} & \multirow{3}{*} {{\tiny 9539}}\\
{\tiny AUG4} & {\tiny 7430} & {\tiny 7813}&&&\\ 
{\tiny SEP4} & {\tiny 7422} & {\tiny 8100} &&&\\ \hline

{\tiny JUL2}& {\tiny 287906} & {\tiny 287906} &\multirow{3}{*}{{\tiny 170002}} & \multirow{3}{*}{{\tiny 442208}} & \multirow{3}{*} {{\tiny 47944}}\\ 
{\tiny AUG2}  & {\tiny 328761} & {\tiny 339059} & & & \\ 
{\tiny SEP2}& {\tiny 332398} &{\tiny 354572}& & &\\ \hline

%& {\small \#soci links} & {\small \#all soci links} & {\small \#soci nodes}& {\small\#pos attri links} &{\small\#attri nodes}\\ \hline
%
%{\small JUL4} & {\small 7062}& {\small 7062} &\multirow{3}{*}{{\small 5200}} & \multirow{3}{*}{{\small 24690}} & \multirow{3}{*} {{\small 9539}}\\
%{\small AUG4} & {\small 7430} & {\small 7813}&&&\\ 
%{\small SEP4} & {\small 7422} & {\small 8100} &&&\\ \hline
%
%{\small JUL2}& {\small 287906} & {\small 287906} &\multirow{3}{*}{{\small 170002}} & \multirow{3}{*}{{\small 442208}} & \multirow{3}{*} {{\small 47944}}\\ 
%{\small AUG2}  & {\small 328761} & {\small 339059} & & & \\ 
%{\small SEP2}& {\small 332398} &{\small 354572}& & &\\ \hline

%{\tiny JUL0 } & {\tiny 6776118}& \multirow{3}{*}{{\tiny 2583198}} &  \multirow{3}{*} {{\tiny 64015}}\\
%{\tiny AUG0}  & {\tiny 8544876} & & \\ 
%{\tiny SEP0 } & {\tiny 8816930} & & \\ \hline
\end{tabular}
\label{table:net_stat_a}
%} 
%\hspace{0.0cm}
%\subfloat[]{
%\centering
%\addtolength{\tabcolsep}{-5pt}
%\begin{tabular}{|c|c|c|c|c|} \hline
%& \#soci links & \#soci nodes& \#attri links &\#attri nodes\\ \hline
%
%
%JUL4 &7062& \multirow{3}{*}{5200} & \multirow{3}{*}{24690} &  \multirow{3}{*} {9539}\\
%AUG4 &7813&& &\\ 
%SEP4  & 8100 && &\\ \hline
%
%JUL2& 287906 & \multirow{3}{*}{170002} & \multirow{3}{*}{442208} &  \multirow{3}{*} {47944}\\ 
%AUG2  & 339059 & & & \\ 
%SEP2& 354572& & &\\ \hline
%
%%{\tiny JUL0 } & {\tiny 6776118}& \multirow{3}{*}{{\tiny 2583198}} &  \multirow{3}{*} {{\tiny 64015}}\\
%%{\tiny AUG0}  & {\tiny 8839195} & & \\ 
%%{\tiny SEP0 } & {\tiny 9495886} & & \\ \hline
%\end{tabular}
%\label{table:net_stat_b}
%}
\label{table:net_stat}
\vspace{-4mm}
\end{table}

\section{Experiments}

\begin{table*}[ht]\renewcommand{\arraystretch}{0.6}
\centering
\captionsetup{font=small,labelfont=bf}
\caption{\bf Results for predicting new links. (a)AUC of hop-2 new links on the
train-test pair AUG4-SEP4.  (b)AUC of hop-2 new links on the train-test pair
AUG2-SEP2.  (c) (d) AUC of any hop new links on the train-test pair AUG4-SEP4. The numbers in parentheses are standard deviations.}
\hspace{-1.5cm}
\subfloat[]{
\centering
\addtolength{\tabcolsep}{-5pt}
\begin{tabular}{|c|c|c|} \hline 
{\small Alg} & {\small w/o Attri} & {\small With Attri}\\ \hline
{\small Random}& {\small 0.5000} & {\small 0.5000} \\ \hline
{\small CN-SAN}& {\small 0.6730} & {\small 0.7315} \\ \hline
{\small AA-SAN} & {\small 0.7109} & {\small 0.7476}\\ \hline
{\small LRA-SAN} &{\small 0.6003} & {\small 0.6262} \\ \hline
{\small CN+LRA-SAN} & {\small  0.6969} &{\small \textbf{0.7671}} \\ \hline
{\small AA+LRA-SAN} & {\small 0.7118} & {\small 0.7471}\\ \hline
{\small RWwR-SAN} & {\small 0.6033} & {\small 0.6143}\\ \hline

%{\tiny Alg} & {\tiny w/o Attri} & {\tiny With Attri}\\ \hline
%{\tiny Random}& {\tiny 0.5000} & {\tiny 0.5000} \\ \hline
%{\tiny CN-SAN}& {\tiny 0.6730} & {\tiny 0.7315} \\ \hline
%{\tiny AA-SAN} & {\tiny 0.7109} & {\tiny 0.7476}\\ \hline
%{\tiny LRA-SAN} &{\tiny 0.6003} & {\tiny 0.6262} \\ \hline
%{\tiny CN+LRA-SAN} & {\tiny  0.6969} &{\tiny \textbf{0.7671}} \\ \hline
%{\tiny AA+LRA-SAN} & {\tiny 0.7118} & {\tiny 0.7471}\\ \hline
%{\tiny RWwR-SAN} & {\tiny 0.6033} & {\tiny 0.6143}\\ \hline

%SRW &   &0.6023 \\ \hline
\end{tabular}
\label{table:link-pre-4}
} 
\hspace{-0.3cm}
\subfloat[]{
%\centering
\addtolength{\tabcolsep}{-5pt}
\begin{tabular}{|c|c|c|} \hline 
{\small Alg} & {\small w/o Attri} & {\small With Attri}\\ \hline
{\small Random}& {\small 0.5000}  &	{\small 0.5000}\\ \hline
{\small CN-SAN} & {\small 0.6936} & {\small 0.7508} \\ \hline
{\small AA-SAN} & {\small 0.7638} & {\small \textbf{0.7895}} \\ \hline
{\small LRA-SAN} & {\small 0.6410} & {\small 0.6385}\\ \hline
{\small CN+LRA-SAN} & {\small 0.5642} & {\small 0.6373}\\ \hline
{\small AA+LRA-SAN} &{\small 0.6032} & {\small 0.6557}\\ \hline
{\small RWwR-SAN} &{\small 0.6788}&{\small 0.6912} \\ \hline
%SRW&&0.6786 \\ \hline

%{\tiny Alg} & {\tiny w/o Attri} & {\tiny With Attri}\\ \hline
%{\tiny Random}& {\tiny 0.5000}  &	{\tiny 0.5000}\\ \hline
%{\tiny CN-SAN} & {\tiny 0.6936} & {\tiny 0.7508} \\ \hline
%{\tiny AA-SAN} & {\tiny 0.7638} & {\tiny \textbf{0.7895}} \\ \hline
%{\tiny LRA-SAN} & {\tiny 0.6410} & {\tiny 0.6385}\\ \hline
%{\tiny CN+LRA-SAN} & {\tiny 0.5642} & {\tiny 0.6373}\\ \hline
%{\tiny AA+LRA-SAN} &{\tiny 0.6032} & {\tiny 0.6557}\\ \hline
%{\tiny RWwR-SAN} &{\tiny 0.6788}&{\tiny 0.6912} \\ \hline

\end{tabular}
\label{table:link-pre-2}
}
%\hspace{0.0cm}
%\subfloat[]{
%\centering
%%\small\addtolength{\tabcolsep}{-5pt}
%\begin{tabular}{|c|c|c|c|} \hline 
%Alg & w/o Attri & With Attri \\ \hline
% Random &  0.5000 & 0.5000\\ \hline
%CN-SAN& 0.7406&0.7444\\ \hline
%AA-SAN& 0.8289 &0.8299\\ \hline
%%LRA&  &&\\ \hline
%%CN+LRA&  &&\\ \hline
%%AA+LRA& &&\\ \hline
%\end{tabular}
%\label{table:link-pre-0}
%}
\hspace{-0.3cm}
\subfloat[]{
\centering
\addtolength{\tabcolsep}{-5pt}
\begin{tabular}{|c|c|c|} \hline 
{\small Alg} & {\small w/o Attri} & {\small With Attri}\\ \hline
{\small Random}&{\small 0.5000}&{\small 0.5000} \\ \hline
{\small CN-SAN}&{\small 0.7482}&{\small 0.8298}\\ \hline
{\small AA-SAN} &{\small 0.7483}  &{\small 0.8324} \\ \hline
{\small LRA-SAN} & {\small 0.8075}&{\small 0.8237}\\ \hline
{\small CN+LRA-SAN} & {\small 0.7857} & {\small 0.8651}\\ \hline
{\small AA+LRA-SAN} & {\small 0.8193}&{\small 0.8552} \\ \hline
{\small RWwR-SAN} &{\small 0.9363} &{\small \textbf{0.9548}} \\ \hline

%{\tiny Alg} & {\tiny w/o Attri} & {\tiny With Attri}\\ \hline
%{\tiny Random}&{\tiny 0.5000}&{\tiny 0.5000} \\ \hline
%{\tiny CN-SAN}&{\tiny 0.7482}&{\tiny 0.8298}\\ \hline
%{\tiny AA-SAN} &{\tiny 0.7483}  &{\tiny 0.8324} \\ \hline
%{\tiny LRA-SAN} & {\tiny 0.8075}&{\tiny 0.8237}\\ \hline
%{\tiny CN+LRA-SAN} & {\tiny 0.7857} & {\tiny 0.8651}\\ \hline
%{\tiny AA+LRA-SAN} & {\tiny 0.8193}&{\tiny 0.8552} \\ \hline
%{\tiny RWwR-SAN} &{\tiny 0.9363} &{\tiny \textbf{0.9548}} \\ \hline
\end{tabular}
\label{table:link-pre-4-any-hop}
}
\hspace{-0.3cm}
\subfloat[]{
\centering
\addtolength{\tabcolsep}{-5pt}
\begin{tabular}{|c|c|} \hline 
%{\tiny Alg} & {\tiny AUC}\\ \hline
%{\tiny SLP-I}&{\tiny 0.9128(0.0140)}\\ \hline
%{\tiny SLP-II} &{\tiny 0.9580(0.0017)}  \\ \hline
%{\tiny SLP-SAN-III} & {\tiny 0.9450(0.0007)}\\ \hline
%{\tiny SLP-SAN-VI} & {\tiny \textbf{0.9706(0.0004)}}\\ \hline
%{\tiny SRW} &{\tiny \textbf{0.9383}} \\ \hline

{\small Alg} & {\small AUC}\\ \hline
{\small SLP-I}&{\small 0.9128(0.0140)}\\ \hline
{\small SLP-II} &{\small 0.9580(0.0017)}  \\ \hline
{\small SLP-SAN-III} & {\small 0.9450(0.0007)}\\ \hline
{\small SLP-SAN-VI} & {\small \textbf{0.9706(0.0004)}}\\ \hline
{\small SRW} &{\small 0.9383} \\ \hline

\end{tabular}
\label{table:link-pre-svm}
}
\hspace{-1.2cm}
\label{table:link-pre}
\vspace{-6 mm}
\end{table*}

\subsection{Experimental Setup}
\label{ss:exp_setup}
In our experiments, the main metric used is AUC, Area Under the Receiver
Operating Characteristic (ROC) Curve, which is widely used in the machine
learning and social network communities \cite{infer-missing-link-08, srw11}.
AUC is computed in the manner described in \cite{auc01}, in which both positive
and negative examples are required. In principle, we could use new links or
missing links as positive examples and all non-existing links as negative
examples. However, large-scale social networks tend to be very sparse, e.g.,
the average degree is $4.17$ in SEP2, and, as a result, the number of
non-existing links can be enormous, e.g., SEP2 has around $2.9\times10^{10}$
non-existing links. Hence, computing AUC using all non-existing links in
large-scale networks is typically computationally infeasible.  Moreover,  the
majority of new links in typical online social networks close triangles
\cite{hop-2-08,srw11}, i.e., are hop-2 links. For instance, we find that $58\%$
of the newly added links in Google+ are hop-2 links.  We thus evaluate our
large network experiments using hop-2 link data as in \cite{srw11}, i.e., new
or missing hop-2 links are treated as positive examples and non-existing hop-2
links are treated as negative examples. 

In a social-attribute network, there are two categories of hop-2 links: 1)
those with  two endpoints sharing at least one common social node, and 2) those
with two endpoints sharing only common attribute nodes.
Local algorithms applied to the original social network are unable to predict
hop-2 links in the second category. Thus, we evaluate only with respect to
hop-2 links in the first category, so as not to give unfair advantage to
algorithms running on the social-attribute network.  To better understand
whether the AUC performance computed on hop-2 links can be generalized to
performance on any-hop links, we additionally compute AUC using any-hop links
on the smaller \gplus networks. 

%Our choice to evaluate with hop-2 links also allows us to run RWwR for
%large-scale social networks.  As noted by \cite{srw11}, RWwR (and thus RWwR-SAN)
%is computationally inefficient on large-scale networks.  So, for hop-2 links evaluation,  we run RWwR in hop-2 local neighborhoods and we renormalize the stationary probabilities for hop-2 links. However, we run RWwR on the whole network in experiments of any-hop links on the smaller \gplus network.

In general, different nodes and links can have different weights in
social-attribute networks, representing their relative importance in the
network. In all of our experiments in this paper, we set all weights to be one
and leave it for future work to learn weights.

We compare our link prediction algorithms with Supervised Random Walk
(SRW)~\cite{srw11}, which leverages edge attributes, by transforming node
attributes to edge attributes.  Specifically, we compute the number of common
attributes of the two endpoints of each existing link. As in \cite{srw11},
we also use the number of common neighbors as an edge attribute. We adopt
the Wilcoxon-Mann-Whitney (WMW) loss function and logistic edge strength function
in our implementations as recommended in~\cite{srw11}. 

We compare our attribute inference algorithms with two algorithms, BASELINE and
LINK, introduced by Zheleva and Getoor~\cite{Zheleva09}.  Using only node
attributes, BASELINE first computes a marginal attribute distribution and then
uses an attribute's probability as its score. LINK trains a classifier for each
attribute by flattening nodes as the rows of the adjacency matrix of the social
networks.\footnote{\scriptsize The original LINK algorithm \cite{Zheleva09}
trained a distinct classifier for each attribute type. In our setting an
attribute type, (e.g., Education) can have multiple values, so we
train a classifier for each binary attribute value.} Zheleva and
Getoor~\cite{Zheleva09} found that LINK is the best algorithm when group
memberships are not available.

We use SVM as our classifier in all supervised algorithms. For link prediction,
we extract six topological features (CN-SAN, AA-SAN, LRA-SAN,
CN+LRA-SAN, AA+LRA-SAN and RWwR-SAN) from both pure social networks and
social-attribute networks. Hence, SLP-I, SLP-II, SLP-SAN-III and SLP-SAN-VI use 6,
7, 6 and 12 features, respectively. For attribute inference, we extract 9
topological features for each attribute link. We adopt two ranks (detailed
in~\ref{sec:attri-infer}) for each low-rank approximation based algorithms,
thus obtaining 6 features. The other three features are CN-SAN, AA-SAN and
RWwR-SAN. To account for the highly imbalanced class distribution of examples
for supervised link prediction and attribute inference we downsample negative examples so
that we have equal number of positive and negative examples (techniques
proposed in ~\cite{Lichtenwalter10, Doppa10} could be used to further
improve the performance).

We use the pattern \emph{dataset1}-\emph{dataset2} to denote a train-test or train-validation
pair, with \emph{dataset1} a training dataset and \emph{dataset2} a testing or validation
dataset. When conducting experiments to predict new links on the AUG$i$-SEP$i$
train-test pair, SRW, classifiers and hyperparameters of global algorithms, i.e.,
ranks in LRA-SAN, CN+LRA-SAN, and AA+LRA-SAN and the restart probability
$\alpha$ in RWwR-SAN, are learned on the JUL$i$-AUG$i$
train-validation pair. Similarly,  when predicting missing links on train-test
pair AUG$i$-JUL$i$, they are learned on train-validation pair SEP$i$-AUG$i$, where $i=2,4$.

The CN-SAN and AA-SAN algorithms are implemented in Python 2.7 while the
RWwR-SAN algorithm and Supervised Random Walk (SRW) are implemented in Matlab, and all of them are run on a desktop with a 3.06 GHz Intel Core i3 and 4GB of main memory. LRA-SAN, CN+LRA-SAN and
AA+LRA-SAN algorithms are implemented in Matlab and run on an x86-64
architecture using a single 2.60 Ghz core and 30GB of main memory.

\subsection{Experimental Results}
In this section we present evaluations of the algorithms on the Google+
dataset. We first show that incorporating attributes via the SAN model improves
the performance of both unsupervised and supervised link prediction algorithms.
Then we demonstrate that inferring attributes via link prediction algorithms within the SAN model achieves state-of-the-art performance.  
Finally, we show that by combining attribute inference and link prediction in
an iterative fashion, we achieve even greater accuracy on the link prediction
task.

\begin{figure}[!h]
\centering
\includegraphics[width=0.25 \textwidth, height = 1.5 in]{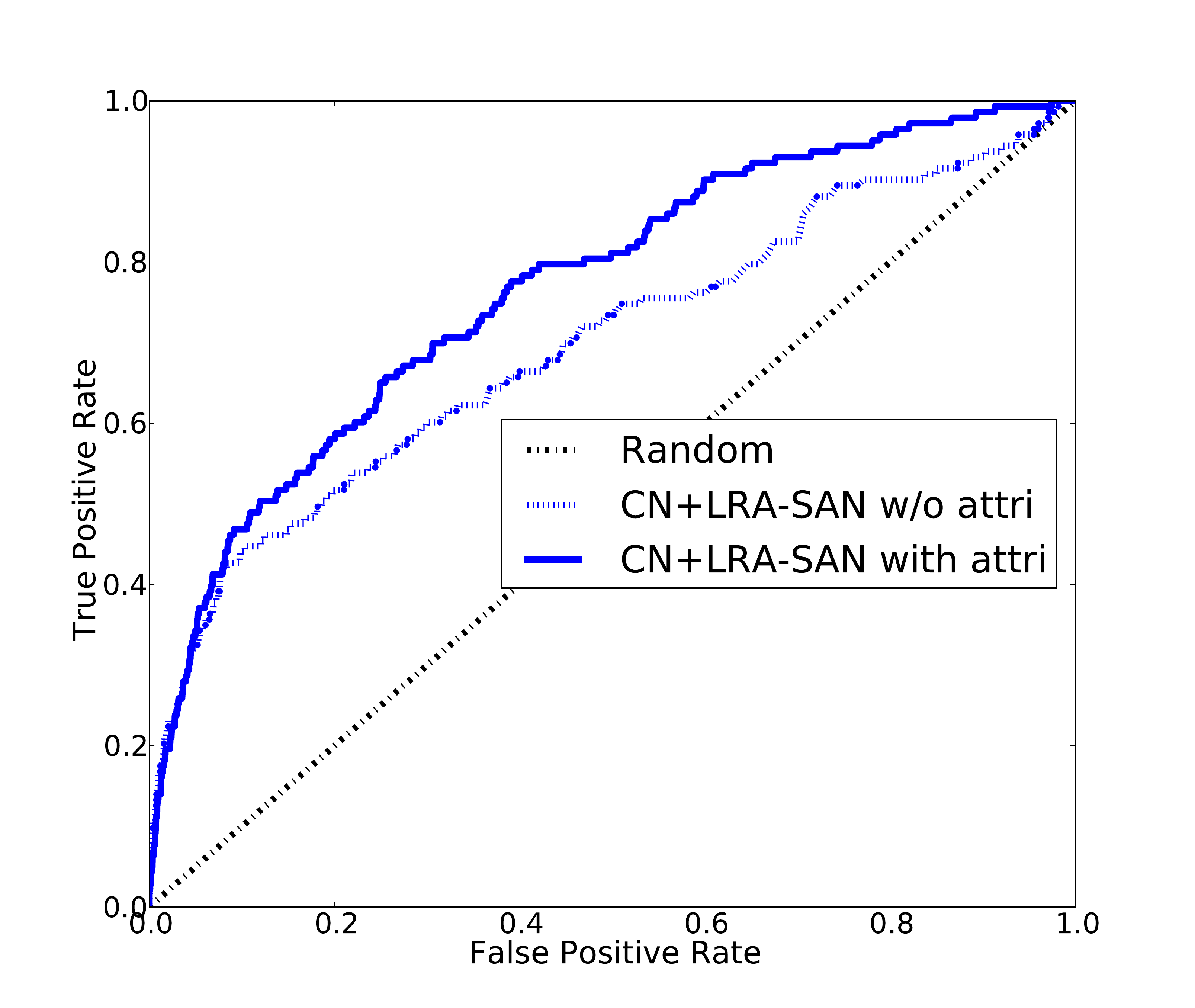}
\captionsetup{font=small,labelfont=bf}
\caption{\bf  ROC curves of the CN+LRA-SAN algorithm for predicting new links.
AUG4-SEP4 is the train-test pair. JUL4-AUG4 is the train-validation pair.}
\label{figure:exp1-link-pre-roc-4}
\vspace{-5 mm}
\end{figure}

\subsubsection{Link Prediction}
\label{sec:link-prediction}
To demonstrate the benefits of combining node attributes and network structure,
we run the SAN-based link prediction algorithms described in
Section~\ref{sec:algorithms} both on the original social networks and on the
corresponding social-attribute networks (recall that the SAN-based unsupervised algorithms
reduce to standard unsupervised link prediction algorithms when working solely with the
original social networks).\\

\noindent
{\bf Predicting New Links} Table \ref{table:link-pre} shows the AUC results of
predicting new links for each of our datasets. We are able to draw a number of
conclusions from these results. First, the SAN model improves every
unsupervised learning algorithm on every dataset, save for LRA-SAN on
AUG2-SEP2. Second, Table~\ref{table:link-pre-svm} shows that attributes
also improve supervised link prediction performance since SLP-SAN-VI,
SLP-SAN-III and SLP-II outperform SLP-I. Moreover, SLP-SAN-VI, which adopts
features extracted from both social networks and social-attribute networks,
achieves the best performance, thus demonstrating the power of the SAN model.
Third, comparing RWwR-SAN in Table~\ref{table:link-pre-4-any-hop} and SRW in
Table~\ref{table:link-pre-svm}, we observe that the SAN model is better
than SRW at leveraging node attributes since RWwR-SAN with attributes outperforms
SRW.  This result is not surprising given that SRW is designed for edge
attributes and when transforming node attributes to edge attributes, we lose
some information. For instance, as illustrated in Fig.~\ref{figure:san}, nodes
$u_2$ and $u_5$ share the attribute San Francisco. When transforming node
attributes to edge attributes, this common attribute information is lost since
$u_2$ and $u_5$ are not linked.
  
Fig.~\ref{figure:exp1-link-pre-roc-4} shows the ROC curves of the CN+LRA-SAN
algorithm. We see that curve of CN+LRA-SAN with attributes dominates that of
CN+LRA-SAN without attributes, demonstrating the power of the SAN model to
effectively incorporate the additional predictive information of attributes. \\

%Second, the local algorithms (CN-SAN and AA-SAN) outperform the pure global
%algorithms (LRA-SAN and RWwR-SAN) under hop-2 link evaluation.  This is likely
%explained by the fact that hop-2 link evaluation is a measure of local
%prediction performance and hence favors methods like CN-SAN and AA-SAN that
%only predict local links.  Indeed, we see in
%Table~\ref{table:link-pre-4-any-hop} that RWwR-SAN outperforms the
%performance of CN-SAN and AA-SAN under an any-hop evaluation.  

%Third, when comparing Table \ref{table:link-pre-4} and \ref{table:link-pre-4-any-hop},
%we find that, for global algorithms, the absolute
%improvement of AUC is similar under hop-2 evaluation and any-hop evaluation.
%However, for CN-SAN and AA-SAN, the AUC improvement under any-hop evaluation is
%significantly higher than that under hop-2 evaluation.  For example, AA-SAN
%achieves a 0.04 AUC improvement under hop-2 evaluation but improves by 0.08
%under any-hop evaluation.  This is unsurprising when we note that the SAN model allows
%local algorithms like CN-SAN and AA-SAN to form more global predictions by
%introducing common attribute neighbors for nodes that were otherwise distant in
%the social network.  This effect is more muted for the global
%algorithms which are able to form global predictions even when attributes are
%not considered.\\

\begin{table*}[]\renewcommand{\arraystretch}{0.5}
\centering
\captionsetup{font=small,labelfont=bf}
\caption{\bf Results for predicting missing links. (a) AUC of hop-2 missing
links on the train-test pair AUG4-JUL4. (b) AUC of hop-2 missing links on the
train-test pair AUG2-JUL2. (c)-(f) AUC of any-hop missing links on the
train-test pair AUG4-JUL4. Missing links in both categories 1 and 2 are used in
(c) and (d). Missing links in Category 1 are used in (e) and (f). 
%For categories of missing links, please refer to the main context. 
The numbers in parentheses are standard deviations.}
% (c) AUC of SVMs for  any-hop missing links on the train-test pair AUG4-JUL4. Any-hop missing links in category 1 and 2 are used. The numbers in parentheses are standard deviations.
%(c) AUC of hop-2 missing links on train-test pair AUG0-JUL0. 70\% nodes in
%this dataset have no observed attributes.
%(d) AUC of any-hop missing links on the train-test pair AUG4-JUL4. Any-hop missing links in category 1 and 2 are used. (e) AUC of any-hop missing links on the train-test pair AUG4-JUL4. Any-hop missing links in category 1 are used. For categories of missing links, please refer to the context. (f)AUC of SVMs for  any-hop missing links on the train-test pair AUG4-JUL4. Any-hop missing links in category 1 are used. For categories of missing links, please refer to the context.}
\subfloat[]{
\centering
\addtolength{\tabcolsep}{-5pt}
\begin{tabular}{|c|c|c|} \hline 
%{\tiny Alg} & {\tiny w/o Attri} & {\tiny With Attri}\\ \hline
%{\tiny Random}&{\tiny 0.5000}&{\tiny 0.5000}\\ \hline
%{\tiny CN-SAN}&{\tiny 0.7180}&{\tiny 0.7925}\\ \hline
%{\tiny AA-SAN}&{\tiny 0.7437}&{\tiny 0.7697}\\ \hline
%{\tiny LRA-SAN}&{\tiny 0.6569}&{\tiny 0.6237}\\ \hline
%{\tiny CN+LRA-SAN}&{\tiny 0.7147}&{\tiny \textbf{0.7986}}\\ \hline
%{\tiny AA+LRA-SAN}&{\tiny 0.7410} & {\tiny 0.7668}\\ \hline
%{\tiny RWwR-SAN}&{\tiny 0.5731}&{\tiny 0.5676}\\ \hline

{\small Alg} & {\small w/o Attri} & {\small With Attri}\\ \hline
{\small Random}&{\small 0.5000}&{\small 0.5000}\\ \hline
{\small CN-SAN}&{\small 0.7180}&{\small 0.7925}\\ \hline
{\small AA-SAN}&{\small 0.7437}&{\small 0.7697}\\ \hline
{\small LRA-SAN}&{\small 0.6569}&{\small 0.6237}\\ \hline
{\small CN+LRA-SAN}&{\small 0.7147}&{\small \textbf{0.7986}}\\ \hline
{\small AA+LRA-SAN}&{\small 0.7410} & {\small 0.7668}\\ \hline
{\small RWwR-SAN}&{\small 0.5731}&{\small 0.5676}\\ \hline

%SRW&&0.5731\\ \hline
\end{tabular}
\label{table:infer-missing-link-4}
}
\hspace{0.3cm}
\subfloat[]{
\centering
\addtolength{\tabcolsep}{-5pt}
\begin{tabular}{|c|c|c|} \hline 
%{\tiny Alg} & {\tiny w/o Attri} & {\tiny With Attri}\\ \hline
%{\tiny Random}&{\tiny 0.5000}&{\tiny 0.5000}\\ \hline
%{\tiny CN-SAN}&{\tiny 0.6938}&{\tiny 0.7309}\\ \hline
%{\tiny AA-SAN}&{\tiny 0.7633}&{\tiny \textbf{0.7796}}\\ \hline
%{\tiny LRA-SAN}&{\tiny 0.6044}&{\tiny 0.6059}\\ \hline
%{\tiny CN+LRA-SAN}&{\tiny 0.5816}&{\tiny 0.6266}\\ \hline
%{\tiny AA+LRA-SAN}&{\tiny 0.6212}&{\tiny 0.6569}\\ \hline
%{\tiny RWwR-SAN}&{\tiny 0.6595} &{\tiny 0.6706}\\ \hline

{\small Alg} & {\small w/o Attri} & {\small With Attri}\\ \hline
{\small Random}&{\small 0.5000}&{\small 0.5000}\\ \hline
{\small CN-SAN}&{\small 0.6938}&{\small 0.7309}\\ \hline
{\small AA-SAN}&{\small 0.7633}&{\small \textbf{0.7796}}\\ \hline
{\small LRA-SAN}&{\small 0.6044}&{\small 0.6059}\\ \hline
{\small CN+LRA-SAN}&{\small 0.5816}&{\small 0.6266}\\ \hline
{\small AA+LRA-SAN}&{\small 0.6212}&{\small 0.6569}\\ \hline
{\small RWwR-SAN}&{\small 0.6595} &{\small 0.6706}\\ \hline

%SRW&&0.6597\\ \hline
\end{tabular}
\label{table:infer-missing-link-2}
}  
\hspace{0.3cm}
\subfloat[]{
\centering
\addtolength{\tabcolsep}{-5pt}
\begin{tabular}{|c|c|} \hline 
%{\tiny Alg} & {\tiny AUC}\\ \hline
%{\tiny SLP-I}&{\tiny 0.5453(0.0120)}\\ \hline
%{\tiny SLP-II} &{\tiny 0.6991(0.0065)}  \\ \hline
%{\tiny SLP-SAN-III} & {\tiny 0.7161(0.0030)}\\ \hline
%{\tiny SLP-SAN-VI} & {\tiny \textbf{0.8481(0.0022)}}\\ \hline

{\small Alg} & {\small AUC}\\ \hline
{\small SLP-I}&{\small 0.5453(0.0120)}\\ \hline
{\small SLP-II} &{\small 0.6991(0.0065)}  \\ \hline
{\small SLP-SAN-III} & {\small 0.7161(0.0030)}\\ \hline
{\small SLP-SAN-VI} & {\small \textbf{0.8481(0.0022)}}\\ \hline

%{\tiny SRW} &{\tiny \textbf{0.}} \\ \hline
\end{tabular}
\label{table:infer-missing-link-svm}
}
 
%\hspace{0.5cm}
%\subfloat[]{
%\centering
%%\small\addtolength{\tabcolsep}{-5pt}
%\begin{tabular}{|c|c|c|c|} \hline 
%Alg & w/o Attri&With Attri\\ \hline
%Random & 0.5000 &0.5000 \\ \hline
%CN-SAN&0.7599&0.7630\\ \hline
%AA-SAN&0.8188&0.8197\\ \hline
%%LRA&&\\ \hline
%%CN+LRA&&\\ \hline
%%AA+LRA&&\\ \hline
%\end{tabular}
%\label{table:infer-missing-link-0}
%}
\hspace{0.0cm}
\subfloat[]{
\centering
\addtolength{\tabcolsep}{-5pt}
\begin{tabular}{|c|c|c|} \hline 
%{\tiny Alg} & {\tiny w/o Attri} & {\tiny With Attri}\\ \hline
%{\tiny Random}&{\tiny 0.5000}&{\tiny 0.5000} \\ \hline
%{\tiny CN-SAN}&{\tiny 0.5460}&{\tiny 0.7012}\\ \hline
%{\tiny AA-SAN} &{\tiny 0.5460}&{\tiny 0.7033} \\ \hline
%{\tiny LRA-SAN} &{\tiny 0.5495}&{\tiny 0.6177}\\ \hline
%{\tiny CN+LRA-SAN} & {\tiny 0.5547} &{\tiny 0.7048} \\ \hline
%{\tiny AA+LRA-SAN} & {\tiny 0.5640}& {\tiny 0.7325} \\ \hline
%{\tiny RWwR-SAN}&{\tiny 0.2000}&{\tiny \textbf{0.7619}}\\ \hline

{\small Alg} & {\small w/o Attri} & {\small With Attri}\\ \hline
{\small Random}&{\small 0.5000}&{\small 0.5000} \\ \hline
{\small CN-SAN}&{\small 0.5460}&{\small 0.7012}\\ \hline
{\small AA-SAN} &{\small 0.5460}&{\small 0.7033} \\ \hline
{\small LRA-SAN} &{\small 0.5495}&{\small 0.6177}\\ \hline
{\small CN+LRA-SAN} & {\small 0.5547} &{\small 0.7048} \\ \hline
{\small AA+LRA-SAN} & {\small 0.5640}& {\small 0.7325} \\ \hline
{\small RWwR-SAN}&{\small 0.2000}&{\small \textbf{0.7619}}\\ \hline

\end{tabular}
\label{table:infer-missing-link-4-any-hop}
}
\hspace{0.3cm}
\subfloat[]{
\centering
\addtolength{\tabcolsep}{-5pt}
\begin{tabular}{|c|c|c|} \hline 
%{\tiny Alg} & {\tiny w/o Attri} & {\tiny With Attri}\\ \hline
%{\tiny Random}&{\tiny 0.5000}&{\tiny 0.5000} \\ \hline
%{\tiny CN-SAN}&{\tiny 0.7329}&{\tiny 0.7765}\\ \hline
%{\tiny AA-SAN} &{\tiny 0.7330}&{\tiny 0.7784} \\ \hline
%{\tiny LRA-SAN} &{\tiny 0.7316}&{\tiny 0.7401}\\ \hline
%{\tiny CN+LRA-SAN} & {\tiny 0.7515} &{\tiny 0.7510} \\ \hline
%{\tiny AA+LRA-SAN} & {\tiny 0.8104}& {\tiny 0.8116} \\ \hline
%{\tiny RWwR-SAN}&{\tiny 0.7797}&{\tiny \textbf{0.8838}}\\ \hline

{\small Alg} & {\small w/o Attri} & {\small With Attri}\\ \hline
{\small Random}&{\small 0.5000}&{\small 0.5000} \\ \hline
{\small CN-SAN}&{\small 0.7329}&{\small 0.7765}\\ \hline
{\small AA-SAN} &{\small 0.7330}&{\small 0.7784} \\ \hline
{\small LRA-SAN} &{\small 0.7316}&{\small 0.7401}\\ \hline
{\small CN+LRA-SAN} & {\small 0.7515} &{\small 0.7510} \\ \hline
{\small AA+LRA-SAN} & {\small 0.8104}& {\small 0.8116} \\ \hline
{\small RWwR-SAN}&{\small 0.7797}&{\small \textbf{0.8838}}\\ \hline

\end{tabular}
\label{table:infer-missing-link-4-any-hop-1}
}
\hspace{0.3cm}
\subfloat[]{
\centering
\addtolength{\tabcolsep}{-5pt}
\begin{tabular}{|c|c|} \hline 
%{\tiny Alg} & {\tiny AUC}\\ \hline
%{\tiny SLP-I}&{\tiny 0.8023(0.0088)}\\ \hline
%{\tiny SLP-II} &{\tiny 0.8403(0.0033)}  \\ \hline
%{\tiny SLP-SAN-III} & {\tiny 0.8620(0.0080)}\\ \hline
%{\tiny SLP-SAN-VI} & {\tiny \textbf{0.8854(0.0324)}}\\ \hline

{\small Alg} & {\small AUC}\\ \hline
{\small SLP-I}&{\small 0.8023(0.0088)}\\ \hline
{\small SLP-II} &{\small 0.8403(0.0033)}  \\ \hline
{\small SLP-SAN-III} & {\small 0.8620(0.0080)}\\ \hline
{\small SLP-SAN-VI} & {\small \textbf{0.8854(0.0324)}}\\ \hline

%{\tiny SRW} &{\tiny \textbf{0.}} \\ \hline
\end{tabular}
\label{table:infer-missing-link-svm-1}
}
\label{table:infer-link}
\vspace{-6 mm}
\end{table*}

\begin{figure}
 \centering
\subfloat[]{\includegraphics[width=0.25\textwidth, height=1.8in]{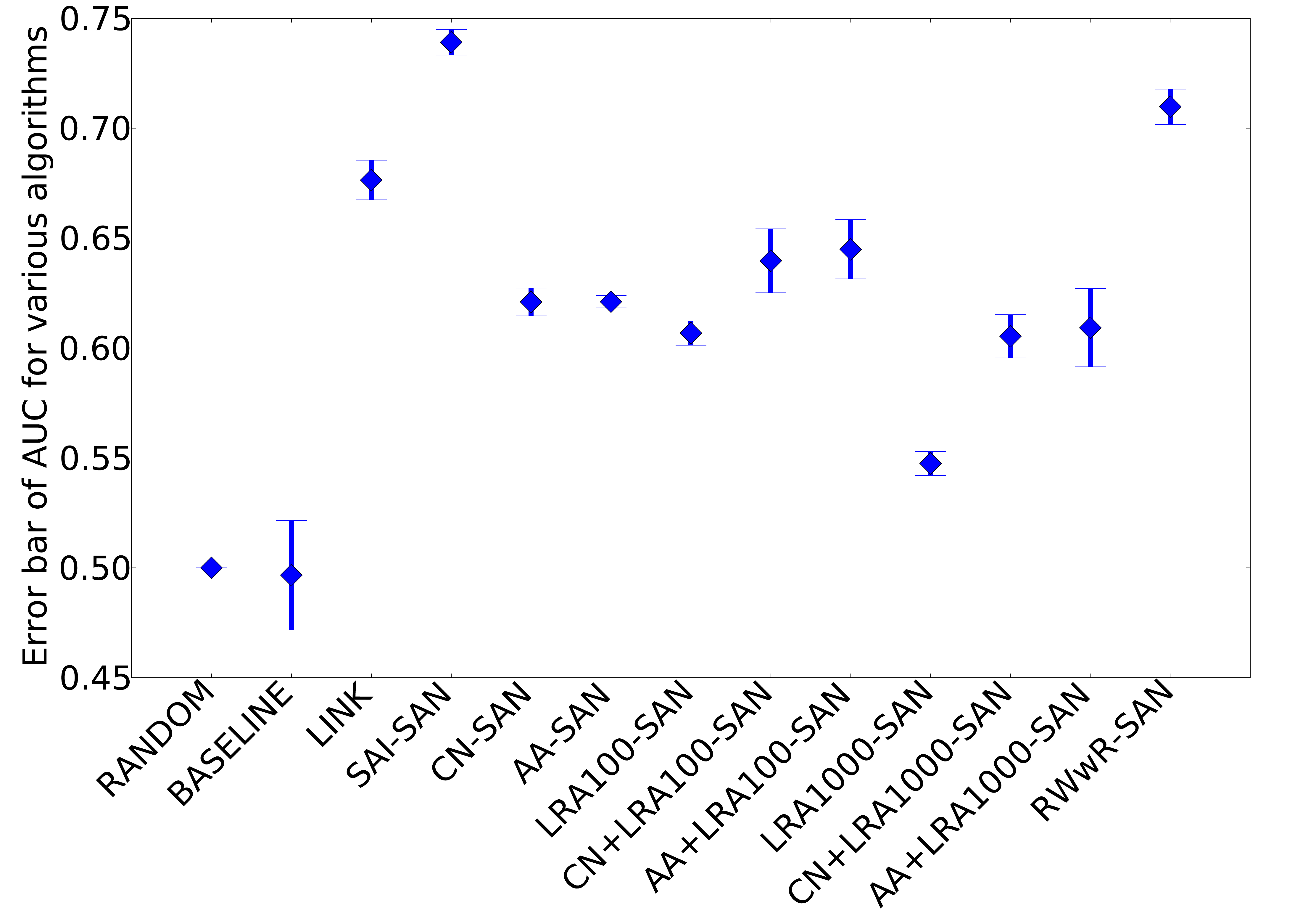} \label{figure:infer-attri-a}}
%\subfloat[]{\includegraphics[width=0.15 \textwidth, height=1.8in]{exp2-b.pdf} \label{figure:infer-attri-b}}
\subfloat[]{\includegraphics[width=0.25\textwidth, height=1.8in]{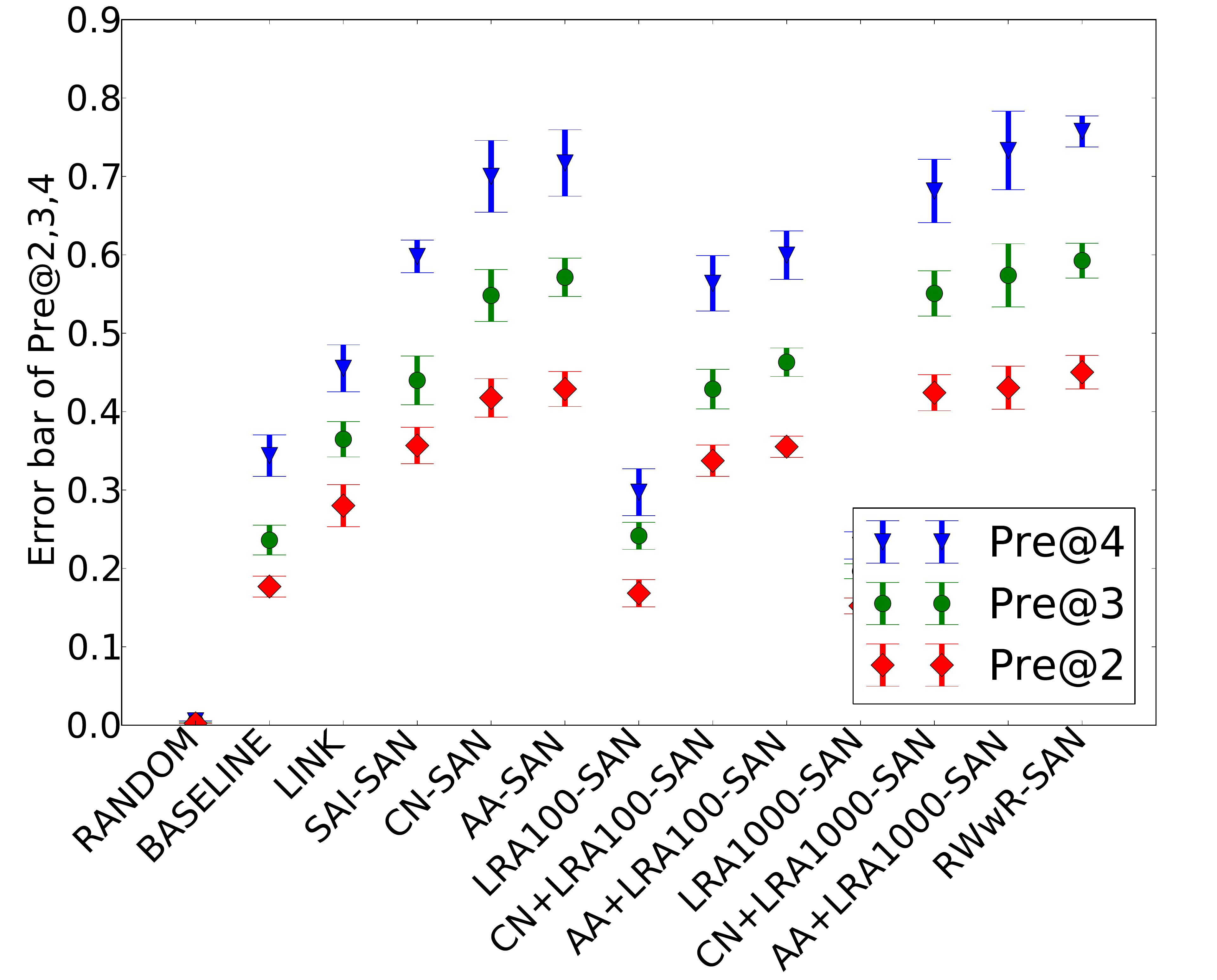} \label{figure:infer-attri-c}}
%\subfloat[]{\includegraphics[width=0.15 \textwidth, height=1.8in]{exp2-d.pdf} \label{figure:infer-attri-d}}
\captionsetup{font=small,labelfont=bf}         
\caption{\bf Performance of various algorithms on attribute inference
on SEP4. (a) AUC under ROC curves. (b) Pre@2,3,4. }
%(c) AUC under ROC curves of RWwR-SAN as a function of the restart probability. (d) Pre@2,3,4 of RWwR-SAN as a function of the restart probability.}
\label{figure:infer-attri}
\vspace{-5 mm}
\end{figure}

\noindent
{\bf Predicting Missing Links} Missing links can be divided into two
categories: 1) links whose two endpoints have some social links in
the training dataset. 2) links with at least one endpoint that has no
social links in the training dataset. Category 1 corresponds to the scenarios
where users block users or users set a part of their friend lists (e.g. family
circles) to be private. Category 2 corresponds to the scenario in which users
hide their entire friend lists. Note that all hop-2 missing links belong to
Category 1. In addition to performing experiments to show that the SAN model improves
missing link prediction, we also perform experiments to explore which category
of missing links is easier to predict. Table \ref{table:infer-link} shows the
results of predicting missing links on various datasets.  As in the new-link
prediction setting, the performance of every algorithm is improved by the SAN
model, except for LRA-SAN on AUG4-JUL4 and RWwR-SAN on AUG4-JUL4 for hop-2
missing links.  

%Interestingly, the
%combination of low-rank approximation and CN-SAN/AA-SAN achieves better AUC on
%AUG4-JUL4 for both hop-2 and any-hop missing links than any other algorithm,
%and the local algorithms perform better than the other algorithms on AUG2-JUL2
%under hop-2 evaluation.  These results are consistent with the new-link
%prediction results.

%When comparing Table \ref{table:infer-missing-link-4} and
%\ref{table:infer-missing-link-4-any-hop}, we observe interesting
%trends.  First, LRA-SAN without attributes works better for hop-2 links, but
%LRA-SAN with attributes works better for any-hop links. Second, for all
%algorithms, AUC improvements of any-hop links are much higher than those of
%hop-2 links, but the absolute AUC values of any-hop links are significantly
%lower than those of hop-2 links. This finding may be explained by the fact that
%some users set their whole friend lists to be publicly invisible. In this case,
%we have no neighborhood information for these users. Therefore their missing
%links are very difficult to predict. 

When comparing Tables~\ref{table:infer-missing-link-4-any-hop}
and~\ref{table:infer-missing-link-4-any-hop-1} or
Tables~\ref{table:infer-missing-link-svm}
and~\ref{table:infer-missing-link-svm-1}, we conclude that the missing links in
Category 2 are harder to predict than those in Category 1.  RWwR-SAN without
attributes performs poorly when predicting any-hop missing links in both
categories (as indicated by the entry with $0.2000$ in
Table~\ref{table:infer-missing-link-4-any-hop}). This poor performance is due
to the fact that RWwR-SAN without attributes assigns zero scores for all the
missing links in Category 2 (positive examples) and positive scores for most
non-existing links (negative examples), making many negative examples rank
higher than positive examples and resulting in a very low AUC.

\subsubsection{Attribute Inference}
\label{sec:attri-infer}
In this section, we focus on inferring attributes using the SAN
model. In our next set of experiments in Section \ref{sec:iterate}, we use the results of
these attribute inference algorithms to further improve link prediction, and
the results of this iterative approach further validate the performance of the
SAN model for attribute inference.  Since the first step of iterative approach of Section
\ref{sec:iterate} involves inferring the top attributes for each node, we
employ an additional performance metric called Pre@$K$ in our attribute
inference experiments.  Compared to AUC, Pre@$K$ better captures the quality of
the top attribute predictions for each user.  Specifically, for each sampled
user, the top-$K$ predicted attributes are selected, and (unnormalized) Pre@$K$
is then defined as the number of positive attributes selected divided by the
number of sampled users. We address score ties in the manner described
in~\cite{McSherry08}. Since most \gplus\ users have a small number of
attributes, we set $K = 2,3,4$ in our experiments. 

When evaluating algorithms for the inference of missing attributes, we require
ground truth data. In general, ground truth for node attributes is difficult to
obtain since it is often not possible to distinguish between negative and
missing attributes.  However, for most users the number of attributes is quite
small, and so we assume that users with many positive attributes have no
missing attributes.  Hence, we evaluate attribute inference on users
that have at least 4 specified attributes, i.e., we work with users in SEP4 and
assume that each attribute link in SEP4 is either positive or negative.

In our experiment, we sample 10\% of the users in SEP4 uniformly at random,
remove their attribute links from SEP4, and evaluate the accuracy with which we
can infer these users' attributes.  All removed positive attribute links are viewed as
positive examples, while all the negative attribute links of the
sampled users are treated as negative examples. We run a variety of algorithms
for attribute inference, and for each algorithm we average the results over 10
random trials. As noted above, we evaluate the performance
of attribute inference using both AUC and Pre@$K$.  

For the low-rank approximation based algorithms, i.e., LRA-SAN, CN+LRA-SAN and
AA+LRA-SAN, we report results using two different ranks, 100 and 1000, and
indicate which was used by the number following the algorithm name in
Fig.~\ref{figure:infer-attri}. We choose these two small ranks for
computational reasons and also based on the fact that low-rank approximation
methods assume that a small number of latent factors (approximately)
describe the social-attribute networks. For RWwR-SAN, we set the restart
probability $\alpha$ to be 0.7.\footnote{{\scriptsize We find that RWwR-SAN performs
consistently across different restart probabilities (results omitted due to 
space constraints).}}

Fig.~\ref{figure:infer-attri} shows the attribute inference results for various
algorithms. Several interesting observations can be made from this figure.
First, under both metrics, all SAN-based algorithms perform better than
BASELINE, save LRA100-SAN and LRA1000-SAN under Pre@2,3,4 metric, which
indicates that the SAN model is good at leveraging network structure to infer
missing attributes.  
%Second, we find that AUC and Pre@$K$ provide inconsistent conclusions about
%relative algorithm performance.  When comparing Fig.~\ref{figure:infer-attri-a}
%to Fig.~\ref{figure:infer-attri-c}, we find that the mean AUC values suggest
%that low-rank approximation of rank 100 combined with CN-SAN or AA-SAN
%outperforms CN-SAN or AA-SAN alone.  However, the result is reversed for mean
%values of Pre@2,3,4, under which CN-SAN or AA-SAN significantly outperform
%CN+LRA-SAN100 and AA+LRA-SAN100.  Similarly, CN+LRA and AA+LRA with rank 100
%perform better than those with rank 1000 in terms of AUC  but perform worse
%than those with rank 1000 in terms of Pre@2,3,4.  All algorithms have similar
%standard deviation for Pre@2,3,4, but, for AUC, low-rank approximation based
%algorithms have much higher standard deviation, indicating a performance that
%depends more significantly on whose attributes are being inferred.  The
%inconsistencies between the two metrics are expected, since AUC is a global
%measurement while Pre@$K$ is a local one.  
Second, we find that AUC and Pre@$K$ provide inconsistent conclusions about
relative algorithm performance.  For instance, the mean AUC values suggest that
SAI-SAN beats all other algorithms. However, several unsupervised algorithms
outperform SAI-SAN with respect to Pre@2,3,4. The inconsistencies between the
two metrics are expected since AUC is a global measurement while Pre@$K$ is a
local one.  Our SAI-SAN algorithm dominates LINK under both AUC and Pre@2,3,4
metrics, thus demonstrating the power of mapping attribute inference to link
prediction with the SAN model.

%Third, RWwR-SAN achieves the highest mean AUC and Pre@2,3,4 for inferring
%attributes, even better than SAI-SAN! This could be explained by the fact that SAI-SAN flattens the adjacency matrix, thus losing rich structural information.
 
%Recall
%that, for link prediction, CN-SAN and AA-SAN always perform better than
%RWwR-SAN, and all the other global algorithms perform better than RWwR-SAN on
%the smaller datasets AUG4-SEP4/AUG4-JUL4. This means that RWwR-SAN is better at
%inferring attributes than predicting links.

%\begin{figure*}[!t]
% \centering
% \includegraphics[width=1 \textwidth, height=2.7in]{exp2.pdf}      
%\captionsetup{font=small,labelfont=bf}         
%\caption{\bf Performance of various algorithms on inferring missing attributes
%on SEP4. We randomly sample 10\% of the users of SEP4 and withhold their
%attributes for ground truth evaluation data. Various algorithms are used to
%infer attributes for these sampled users, and results are averaged over 10
%trials.  \emph{Left:} AUC under ROC curves. \emph{Right:} Pre@2,3,4. }
% \label{figure:infer-attri}
%\end{figure*}

\begin{table}[ht]\renewcommand{\arraystretch}{0.6}
\centering
\captionsetup{font=small,labelfont=bf}
\caption{\bf Results for iteratively inferring attributes and predicting links.
(a) on the AUG4-SEP4 train-test pair. (b) on the  AUG4-JUL4 train-test pair.
Results are averaged over 10 trials. The numbers in parentheses are standard
deviations.}
\subfloat[]{
\centering
\addtolength{\tabcolsep}{-5pt}
%\small\addtolength{\tabcolsep}{-5pt}
\begin{tabular}{|c|c|c|c|} \hline 
%{\tiny Alg} & {\tiny w/o Attri} & {\tiny With Attri} & {\tiny With Inferred Attri}\\ \hline
%{\tiny Random} & {\tiny 0.5000(0)}  &  {\tiny 0.5000(0)} & {\tiny 0.5000(0)}\\ \hline
%{\tiny CN-SAN} & {\tiny 0.6730(0)} & {\tiny 0.7174(0.0077)} & {\tiny 0.7291(0.0063)}\\ \hline
%{\tiny AA-SAN} & {\tiny 0.7109(0)} & {\tiny 0.7408(0.0063)} & {\tiny 0.7440(0.0026)}\\ \hline
%{\tiny LRA-SAN}& {\tiny 0.6003(0)} & {\tiny 0.6274(0.0052)} & {\tiny 0.6320(0.0055)}\\ \hline
%{\tiny CN+LRA-SAN} & {\tiny 0.6969(0)} & {\tiny 0.7497(0.0134)} & {\tiny \textbf{0.7534}(0.0084)}\\ \hline
%{\tiny AA+LRA-SAN} & {\tiny 0.7111(0)} & {\tiny 0.7373(0.0050)} & {\tiny 0.7442(0.0032)}\\ \hline

{\small Alg} & {\small w/o Attri} & {\small With Attri} & {\small With Inferred Attri}\\ \hline
{\small Random} & {\small 0.5000(0)}  &  {\small 0.5000(0)} & {\small 0.5000(0)}\\ \hline
{\small CN-SAN} & {\small 0.6730(0)} & {\small 0.7174(0.0077)} & {\small 0.7291(0.0063)}\\ \hline
{\small AA-SAN} & {\small 0.7109(0)} & {\small 0.7408(0.0063)} & {\small 0.7440(0.0026)}\\ \hline
{\small LRA-SAN}& {\small 0.6003(0)} & {\small 0.6274(0.0052)} & {\small 0.6320(0.0055)}\\ \hline
{\small CN+LRA-SAN} & {\small 0.6969(0)} & {\small 0.7497(0.0134)} & {\small \textbf{0.7534}(0.0084)}\\ \hline
{\small AA+LRA-SAN} & {\small 0.7111(0)} & {\small 0.7373(0.0050)} & {\small 0.7442(0.0032)}\\ \hline

\end{tabular}
\label{table:iter-link-pre-a}
\label{table:iter-link-pre}
}

\subfloat[]{
\centering
\addtolength{\tabcolsep}{-5pt} 
%\begin{table}[ht]\renewcommand{\arraystretch}{0.7}
%\centering
%\captionsetup{font=small,labelfont=bf}
%\caption{\bf  AUC Results for iteratively inferring attributes and predicting
%missing links on the  AUG4-JUL4 is the train-test pair. Results are averaged over 10
%trials. The numbers in parentheses are standard deviations.}
%\small\addtolength{\tabcolsep}{-5pt}
\begin{tabular}{|c|c|c|c|} \hline 
%{\tiny Alg} &  {\tiny w/o Attri} & {\tiny With Attri} & {\tiny With Inferred Attri}\\ \hline
%{\tiny Random} & {\tiny 0.5000(0)} & {\tiny 0.5000(0)} & {\tiny 0.5000(0)}\\ \hline
%{\tiny CN-SAN} & {\tiny 0.7180(0)} & {\tiny 0.7780(0.0173)} & {\tiny 0.7856(0.0100)}\\ \hline
%{\tiny AA-SAN} & {\tiny 0.7437(0)} & {\tiny 0.7626(0.0100)} & {\tiny 0.7661(0.0045)}\\ \hline
%{\tiny LRA-SAN} & {\tiny 0.6569(0)} & {\tiny 0.6189(0.0105)} & {\tiny 0.6134(0.0157)}\\ \hline
%{\tiny CN+LRA-SAN} & {\tiny 0.7147(0)} & {\tiny 0.7838(0.0256)} & {\tiny \textbf{0.7969}(0.0059)}\\ \hline
%{\tiny AA+LRA-SAN} & {\tiny 0.7410(0)} & {\tiny 0.7591(0.0118)} & {\tiny 0.7673(0.0051)}\\ \hline

{\small Alg} &  {\small w/o Attri} & {\small With Attri} & {\small With Inferred Attri}\\ \hline
{\small Random} & {\small0.5000(0)} & {\small 0.5000(0)} & {\small 0.5000(0)}\\ \hline
{\small CN-SAN} & {\small 0.7180(0)} & {\small 0.7780(0.0173)} & {\small 0.7856(0.0100)}\\ \hline
{\small AA-SAN} & {\small 0.7437(0)} & {\small 0.7626(0.0100)} & {\small 0.7661(0.0045)}\\ \hline
{\small LRA-SAN} & {\small 0.6569(0)} & {\small 0.6189(0.0105)} & {\small 0.6134(0.0157)}\\ \hline
{\small CN+LRA-SAN} & {\small 0.7147(0)} & {\small 0.7838(0.0256)} & {\small \textbf{0.7969}(0.0059)}\\ \hline
{\small AA+LRA-SAN} & {\small 0.7410(0)} & {\small 0.7591(0.0118)} & {\small 0.7673(0.0051)}\\ \hline

\end{tabular}
\label{table:iter-infer-missing-link-a}
\label{table:iter-infer-missing-link}
}
\vspace{-6 mm}
\end{table}

%\subsubsection{Iterative Attribute and Link Inference}
\subsubsection{Iterative Attribute and Link Inference}
\label{sec:iterate}
Section~\ref{sec:link-prediction} demonstrated that knowledge of a user's
attributes can lead to significant improvements in link prediction.  However,
in real-world social networks like Google+, the vast majority of user
attributes are missing (see Fig. \ref{figure:node-attri}).  To increase the
realized benefits of social-attribute networks with few attributes, we propose
first inferring missing attributes for each user whose attributes are
missing and then performing link prediction on the inferred social-attribute
networks.  Recall that SAI-SAN achieves the best AUC, RWwR-SAN achieves the
best Pre@$K$ in inferring attributes (see Fig.~\ref{figure:infer-attri}) and
AA-SAN achieves comparable Pre@$K$ results while being more scalable.
Thus, in the following experiments, we use AA-SAN to first infer the top-$K$
missing attributes for users, and subsequently perform link prediction using
various methods. 

In our experiments, when we are working on the pair \emph{train-test}, we sample
10\% of the users of \emph{train} uniformly at random and remove their
attributes.  We then run three variants of link prediction algorithms: i)
without attributes, ii) with only the remaining attributes, and iii) with the
remaining attributes along with the inferred attributes.  The top-4 attributes
are inferred for each sampled user by AA-SAN. We report the results averaged
over 10 trials. The hyperparameters of the global algorithms are the same as
those in (Section \ref{sec:link-prediction}), which are learned from the
corresponding train-validation pair.

Table \ref{table:iter-link-pre} shows the results of first inferring attributes
and then predicting new links on the AUG4-SEP4 train-test pair.
Table~\ref{table:iter-infer-missing-link} shows the results of first inferring
attributes and then predicting missing links on the AUG4-JUL4 train-test pair.
We see that the inferred attributes improve the performance of all algorithms
except LRA-SAN on predicting missing links, which is unable to make use of
attributes as demonstrated earlier in Table~\ref{table:infer-missing-link-4}.
The AUCs obtained with inferred attributes for all other algorithms are very
close to those obtained with all positive attributes as shown in Table
\ref{table:link-pre-4}.  This further demonstrates that AA-SAN is an effective
algorithm for attribute inference.

\section{Related Work}
A wide range of link prediction methods have been developed.  
%For instance,
%models of complex networks, such as Preferential Attachment \cite{pa99} and
%Hierarchical model~\cite{infer-missing-link-08} can be viewed as models for
%predicting links.  
%%%Clauset et al.~\cite{infer-missing-link-08} propose a hierarchical model to predict
%%missing links, and Kim and Leskovec~\cite{infer-missing-link11} introduce an
%%approach based on  the Kronecker graphs model \cite{kronecker} to predict both
%%missing nodes and missing links.  
Liben-Nowell and Kleinberg \cite{link-pre-survey03} surveyed a
set of unsupervised link prediction algorithms. Li~\cite{Li11} proposed Maximal
Entropy Random Walk (MERW). Lichtenwalter et al.~\cite{Lichtenwalter10}
proposed the PropFlow algorithm which is similar to RWwR but more localized. However, none of these approaches leverage node attribute information.  

Link prediction methods leveraging attribute information first appear in the
relational learning community \cite{relational-link-pre03, latent-link-pre09,
Bilgic07, Yu06}.  However, these approaches suffer from scalability issues. For
instance, the largest network tested in~\cite{relational-link-pre03} has 
about $3K$ nodes.  Recently, Backstrom and Leskovec \cite{srw11} proposed the Supervised
Random Walk (SRW) algorithm to leverage edge attributes. However, SRW does not handle the scenario in which two nodes share common
attributes (e.g.  nodes $u_2$ and $u_5$ in Fig.~\ref{figure:san}), but no edge
already exists between them. Mapping link prediction to a classification
problem~\cite{Hasan06, Lichtenwalter10, Doppa10} is another way to incorporate
attributes. We have shown that classifiers using features extracted from the
SAN model perform very well. Yang et al.~\cite{HongYang11} proposed to jointly predict links and propagate node interests (e.g., music interest). Their algorithm relies on the assumption that each node interest has a set of explicit attributes. As a result, their algorithm cannot be applied to our scenario in which it's hard (if possible) to extract explicit attributes for our node attributes.   

Previous works in \cite{infer-attri-1, infer-attri-2} aim at
inferring node attributes (e.g., ethnicity and political orientation) using
supervised learning methods with features extracted from user names and
user-generated texts. Zheleva and Getoor~\cite{Zheleva09} map attribute
inference to a relational classification problem. They find that methods using
group information achieve good results. These approaches are complementary to
ours since they use additional information apart from network structure and
node attributes. In this paper, we transform the attribute inference problem
into a link prediction problem with the SAN model. Therefore, any link
prediction algorithm can be used to infer missing attributes. More importantly,
we demonstrate that attribute inference can in turn help link prediction with
the SAN model.
%For a binary attribute $a$, the
%nodes in $\Gamma_{s+}(a)$ and $\Gamma_{s-}(a)$ are taken as positive examples
%and negative examples, respectively. Moreover, we can construct features for
%nodes via flattening the weighted social adjacency matrix $X_S$, i.e. the
%corresponding rows of $X_S$ are taken as the feature vectors of nodes. After
%learning a model, it can then be applied to infer the missing attribute links.
%This flattening idea is also adopted by Zheleva and Getoor~\cite{Zheleva09}.
%Based on the assumption that nodes can only have one attribute value for each
%attribute type (e.g. everyone can only be Female or Male for Gender attribute
%type), Zheleva and Getoor~\cite{Zheleva09} map attribute inference to a
%classification problem and train a classifier for each attribute type. However,
%nodes can have multiple attribute values for some attribute types (e.g. some
%people attend several schools for attribute type School), in which case
%%training a classifier for each attribute type is infeasible. With the SAN model, we can train a binary classifier to each binary attribute.

\section{Conclusion and Future Work}

We comprehensively  evaluate the \emph{Social-Attribute Network} (SAN) model proposed in~\cite{Yin10-1, Yin10} in terms of link prediction and attribute inference. More specifically, we adapt several representative unsupervised and supervised link prediction algorithms to the SAN model to both predict links and infer attributes. Our evaluation with a large-scale novel \gplus
network dataset demonstrates performance improvement for each of these generalized algorithm on both link prediction and attribute inference. Moreover, we demonstrate a further improvement of
link prediction accuracy by using the SAN model in an iterative fashion, first
to infer missing attributes and subsequently to predict links.  Interesting 
avenues for future research include devising an iterative algorithm that
alternates between attribute and link prediction, learning node and edge
weights in the SAN model, 
%for example, the Supervised Random Walk (SRW)
%algorithm \cite{srw11}, 
and incorporating edge attributes, negative node attributes
and mutex edges into large-scale experiments.

\eat{Our SAN model motivates several interesting avenues for future work. First, in
principle, attribute inference and link prediction can be alternated repeatedly
in an iterative fashion. However, the errors of each iteration may compound, so
an effective strategy is required to reduce the negative impact of these
errors.  We view the design of such an iterative algorithm as an interesting
future direction.
Second, the performance of the SAN model could be possibly be improved by
learning node and edge weights for the network, using for example, the
Supervised Random Walk (SRW) algorithm \cite{srw11}.
%method could be combined with the SAN model. Specifically, SRW could be
%potentially used to learn the node and edge weights in the SAN model.  
Third, some supervised link prediction algorithms transform the link
prediction problem into a classification problem using features extracted from
the social network.  These algorithms, e.g., the algorithm presented in
\cite{Hasan06}, could leverage the SAN model to extract features that
incorporate both network structure and attribute information. 
%Third, link prediction algorithms by classification~\cite{Hasan06} could also
%take advantage of the SAN model.  Previous work~\cite{Hasan06} maps link
%prediction to classification problem and extracts features from social
%networks. With the SAN model, we could extract better features.
}

\section{Acknowledgments}
We would like to thank Di Wang, Satish Rao, Mario Frank, Kurt Thomas, and Shobha Venkataraman for insightful feedback. This work is supported by the NSF under Grants No. CCF-0424422, 0311808, 0832943, 0448452, 0842694, 0627511, 0842695, 0808617, 1122732, 0831501 CT-L, by the AFOSR under MURI Award No. FA9550-09-1-0539, by the AFRL under grant No. P010071555, by the Office of Naval Research under MURI Grant No. N000140911081, by the MURI program under AFOSR Grant No. FA9550-08-1-0352, the NSF Graduate Research Fellowship under Grant No. DGE-0946797, the DoD through the NDSEG Program, by Intel through the ISTC for Secure Computing, and by a grant from the Amazon Web Services in Education program. Any opinions, findings, and conclusions or recommendations expressed in this material are those of the author(s) and do not necessarily reflect the views of the funding agencies. 

%\bibliographystyle{abbrv}
%\bibliography{refs}

\end{document}